\documentclass{aa} 

\usepackage{graphicx}
\usepackage{amsmath}
\usepackage{stmaryrd}
\usepackage[dvipsnames]{pstricks}
\usepackage{txfonts}
\usepackage{xcolor}
\usepackage{booktabs}
\usepackage{natbib}
\bibpunct{(}{)}{;}{a}{}{,} 
\usepackage[colorlinks=True, 
            urlcolor=blue,
            linkcolor=blue,
            citecolor= blue,
            dvips]{hyperref}
\usepackage{wasysym}
\newcommand{\od}[1]{\mbox{${\cal O}(#1)$}}
\newcommand{\disp}[1]{\displaystyle #1}

\newcommand{\Amp}{{\cal A}}
\newcommand{\Fbot}{F_{\hbox{\scriptsize bot}}}

\newcommand{\Tbump}{T_{\hbox{\scriptsize bump}}}
\newcommand{\Kmin}{K_{\hbox{\scriptsize min}}}
\newcommand{\Kmax}{K_{\hbox{\scriptsize max}}}
\newcommand{\rhotop}{\rho_{\hbox{\scriptsize top}}}
\newcommand{\Ttop}{T_{\hbox{\scriptsize top}}}

\renewcommand{\na}{ \vec{\nabla} }

\def\Div{\mathop{\hbox{div}}\nolimits}
\newcommand{\lp}{ \left(}
\newcommand{\rp}{ \right)}
\newcommand{\Mach}{\text{Ma}}
\newcommand{\Ft}{{\cal F}_{\text{kin}}}
\newcommand{\Fc}{{\cal F}_{\text{conv}}}
\newcommand{\Fh}{{\cal F}_{\text{ent}}}
\newcommand{\Fr}{{\cal F}_{\text{rad}}}
\newcommand{\Fs}{{\cal F}_{\text{mod}}}

\begin{document}

\title{Convective quenching of stellar pulsations}

\author{T.\ Gastine \and B.\ Dintrans}
\institute{Laboratoire d'Astrophysique de Toulouse-Tarbes, Universit\'e
de Toulouse, CNRS, 14 avenue Edouard Belin, F-31400 Toulouse, France}

\offprints{\href{mailto:thomas.gastine@ast.obs-mip.fr}{thomas.gastine@ast.obs-mip.fr}}

\date{\today,~ accepted for publication}

\abstract
{We study the convection-pulsation coupling that occurs in cold Cepheids close
to the red edge of the classical instability strip. In these stars, the surface
convective zone is supposed to stabilise the radial oscillations excited by the
$\kappa$-mechanism.}
{We study the influence of the convective motions onto the amplitude and the
nonlinear saturation of acoustic modes excited by $\kappa$-mechanism. We are
interested in determining the physical conditions needed to lead to a quenching
of oscillations by convection.}
{We compute two-dimensional nonlinear simulations (DNS) of the
convection-pulsation coupling, in which the oscillations are sustained by a
continuous physical process: the $\kappa$-mechanism. Thanks to both a
frequential analysis and a projection of the physical fields onto an acoustic
subspace, we study how the convective motions affect the unstable radial oscillations.}
{Depending
on the initial physical conditions, two main behaviours are obtained:
(\textit{i}) either the unstable fundamental acoustic mode has a large amplitude, carries
the bulk of the kinetic energy and shows a nonlinear saturation similar to the
purely radiative case; (\textit{ii}) or the convective motions affect
significantly the mode amplitude that remains very weak. In this second case,
convection is quenching the acoustic oscillations. We interpret these
discrepancies in terms of the difference in density contrast: larger
stratification leads to smaller convective plumes that do not affect much the
purely radial modes, while large-scale vortices may quench the
oscillations.}
{}

\keywords{Hydrodynamics - Convection - Instabilities - Stars:
oscillations - Methods: numerical - Stars: Variables: Cepheids}

\maketitle

\section{Introduction}

The cold Cepheids located near the red edge of the classical instability strip
have an important surface convective zone that affects their
pulsation properties \citep[e.g. the reviews
of][]{Gautschy-Saio,buchler-cepheid-review}. The first calculations, done
without any convection-pulsation modelling, predicted red edges much cooler
than the observed ones. In fact, predicting the red edge location requires a
non-adiabatic treatment of this coupling as it was already stated by
\cite{Baker-Kippenhahn}.
 
Several time-dependent convection (TDC) models have been then introduced to
address this coupling \citep{Unno, Gough,  St, K, Xiong89}. More recently, these
different models have been largely improved and have succeeded in reproducing 
the correct location of the red edge,
despite their disagreements with the
physical origin of the mode stabilisation \citep[e.g.][]{Bono, WF,
YKB98,Dupret1, Dupret2}.

However, these models rely on many free parameters \citep[e.g.
the seven dimensionless $\alpha$ coefficients used by][]{YKB98}
that are either fitted to the observations or hardly constrained by theoretical
values such that they give
almost identical results for different parameters sets \citep[see
also][]{Koll02}.
Despite their own limitations (weak density or pressure contrasts, for instance),
2-D and 3-D direct numerical simulations (DNS) are a good way to
investigate the
convection-pulsation coupling as the essential
nonlinearities are fully taken into account. The purpose of this paper is
to present the results of two-dimensional DNS of the
convection-pulsation coupling in which the acoustic oscillations are sustained
in a
self-consistent way by the $\kappa$-mechanism that is responsible for the radial
oscillations of Cepheids \citep{Antares2010}.

In \cite{paperI} and \cite{paperII} (hereafter Papers I and II), we have modelled
the $\kappa$-mechanism in Cepheids by a simplified approach, that is, the
propagation of acoustic waves in a layer of perfect gas in which the
ionisation region is shaped by a hollow in the radiative conductivity
profile. We recall that the instability of Cepheid variables relies on the
blocking of the emerging radiative flux near the opacity bumps that are due to 
the ionisation of light elements like H or He. As the radiative conductivity is proportional to the
inverse of opacity, a conductivity hollow therefore mimics such an opacity
bump. By performing the linear stability of this simplified model, we have 
precisely
determined in Paper I the physical conditions that are required to obtain
unstable radial modes excited by $\kappa$-mechanism:
\textit{(i)}
the conductivity hollow must be sufficiently deep;
\textit{(ii)} this hollow must be also located in a precise zone inside the
star, called ``the
transition region'' \citep{zhevakin,cox-book}, that defines the limit where the
acoustic mode
period $\Pi$ is 
close to the
local thermal timescale $\tau_{\text{therm}}$ as 

\begin{equation}
 \Psi = \dfrac{\tau_{\text{therm}}}{\tau_{\text{dyn}}} =\dfrac{\langle c_v T_0\rangle \Delta m}{\Pi L} \simeq 1,
\label{eq:psi}
\end{equation}
where the $\langle c_v T_0\rangle\Delta m$ term denotes the amount of internal
energy embedded {\it above} the ionisation region and $L$ is the star
luminosity (the ratio between the two defining the thermal time needed
for radiating this internal energy towards the surface).

By starting from the most linearly-unstable setups, we have performed
in Paper II the corresponding nonlinear study by the means of direct
numerical simulations. Thanks to a powerful method
that involves several
projections of the computed fields onto suitable subspaces
\citep[e.g.][]{Bogdan1993}, both the amplitude and energy of each
acoustic mode oscillating in the simulation have been extracted. Their
temporal evolutions
have then emphasised the strong nonlinear coupling existing between the
(unstable) fundamental
acoustic mode and the (stable) second overtone. This preferential coupling relies
on a $2\text{:}1$ resonance as the fundamental period is twice the
second overtone one \citep[e.g.][]{Simon76,Klapp85,Buchler86}. It participates
to the nonlinear saturation of the
system and is also responsible for the well-known ``Hertzsprung's
progression'' observed in the luminosity curves of bump Cepheids (see
Paper II for more details).

Following these first two papers devoted to the purely radiative case, we next
address in this work the possible influence of convection onto the acoustic 
oscillations
excited by $\kappa$-mechanism. We have first
slightly modified the initial physical setup to get a convective zone
that overlaps with the hollow in radiative conductivity.
We have then computed the corresponding 2-D nonlinear simulations and shown 
that a coupling between the convective motions and pulsations
can indeed occur (\S~\ref{sec:setup}).
Thanks to a frequential analysis (\S~\ref{sec:freq}), we have
next emphasised the qualitative discrepancies between two specific
simulations that strongly differ from the role played by convection: in one
simulation, the fundamental acoustic mode excited by $\kappa$-mechanism
seems to be not influenced by the underlying convective motions while
in the other simulation, convection almost cancels it.

In order to measure in more details the acoustic oscillations in these
simulations, we have next applied a projection method similar to the
one used previously in the purely radiative case. That is, we have
determined the contribution of acoustic oscillations and convective motions to
the energy budget (\S~\ref{sec:nrj}). The obtained results show again
two opposite behaviours: {\it
(i)} either the amplitude of the linearly-unstable acoustic mode is large
and the nonlinear saturation seems similar to what was observed in radiative
simulations; {\it (ii)} or the amplitude of pulsations remains very weak
and strongly modulated, while the bulk of kinetic energy lies in convective
motions. In this second case, the convective plumes are quenching the
oscillations in a similar way to what is
expected to occur close to the red edge of the Cepheid instability strip.

The last part of this study focus on the physical conditions
suitable for the quenching of the oscillations by convective motions
(\S~\ref{sec:quench}). We investigate the role of both the convective
flux and the density stratification onto the mode amplitude and we show that a
key parameter for the mode stabilisation could be the density contrast across
the whole layer. Indeed, this stratification has an impact on the typical size
of the convective plumes, as a weak (strong) stratification leads
to large (small) vortices. A clear correlation between the integral
scale of convection and the kinetic energy contained in acoustic modes
is found, that suggests a screening-like effect of the mode pattern by
convective plumes. We finally conclude in \S~\ref{sec:conclu} by
discussing some interesting outlooks of this work in the direction of
time-dependent convection theories.

\section{The model}
\label{sec:setup}

\subsection{The convective setup}
\label{sec:conv}

Our system represents a local zoom around an ionisation region. It is
composed of a 2-D layer ($L_x\times L_z$) filled with a monatomic and perfect
gas with $\gamma = c_p/c_v = 5/3$  and $R^* = c_p-c_v$ the ideal gas
constant ($c_p$ and $c_v$ being the specific heats). As we are dealing with
\emph{local} simulations, the vertical gravity $\vec{g}=-g\vec{e_z}$ and the
kinematic viscosity
$\nu$ are assumed to be constant. Following Papers I and II, the ionisation
region is represented by a temperature-dependent radiative conductivity
profile that mimics an opacity bump:

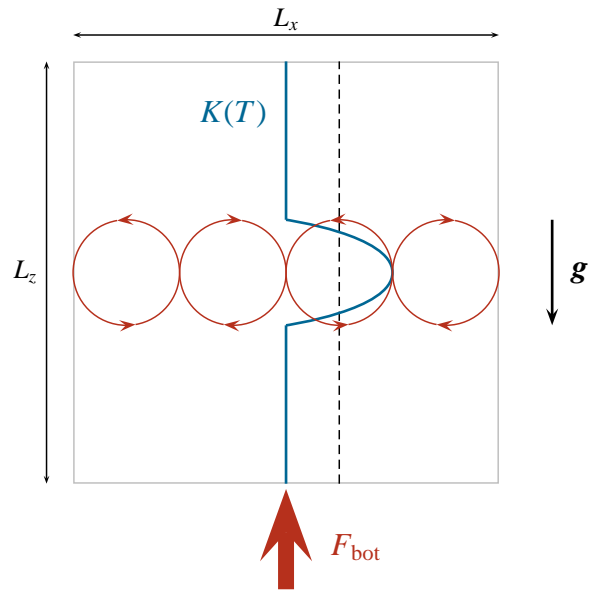
\begin{figure}
 \centering
\begin{pspicture}(0,-1.5)(6,6.5)
\scalebox{0.7}{%
\psframe[linecolor=lightgray](0,0)(8,8)
\psline[linecolor=MidnightBlue,linewidth=0.05](4,0)(4,3)
\psline[linecolor=MidnightBlue,linewidth=0.05](4,5)(4,8)
\rput[c](3,7){\color{MidnightBlue}\LARGE $K(T)$}
\psline[linestyle=dashed](5,0)(5,8)
\pscurve[linecolor=MidnightBlue,linewidth=0.05](4,3)(6,4)(4,5)
\rput(1,4){%
\psarc[linecolor=BrickRed,arrowscale=2]{->}{1}{100}{280}
\psarc[linecolor=BrickRed,arrowscale=2]{->}{1}{280}{100}}
\rput(3,4){%
\psarc[linecolor=BrickRed,arrowscale=2]{<-}{1}{80}{260}
\psarc[linecolor=BrickRed,arrowscale=2]{<-}{1}{260}{80}}
\rput(5,4){%
\psarc[linecolor=BrickRed,arrowscale=2]{->}{1}{100}{280}
\psarc[linecolor=BrickRed,arrowscale=2]{->}{1}{280}{100}}
\rput(7,4){%
\psarc[linecolor=BrickRed,arrowscale=2]{<-}{1}{80}{260}
\psarc[linecolor=BrickRed,arrowscale=2]{<-}{1}{260}{80}}
\psline[linewidth=0.3,linecolor=BrickRed, arrowscale=1.5]{->}(4,-2)(4,-0.1)
\rput[c](5.3,-1.2){\color{BrickRed}\LARGE $F_{\scriptsize\text{bot}}$}
\psline[linewidth=0.05,arrowscale=1.5]{->}(9,5)(9,3)
\rput[c](9.5,4){\LARGE$\vec{g}$}
\psline{<->}(0,8.5)(8,8.5)
\rput[c](4,8.8){\Large $L_x$}
\psline{<->}(-0.5,0)(-0.5,8)
\rput[c](-0.9,4){\Large $L_z$}}
\end{pspicture}
\caption{Sketch of our convective model. The blue curve
corresponds to the radiative conductivity profile, the dashed line is the
Schwarzschild criterion given in Eq.~(\ref{eq:schwarz}), and the large red
arrow represents the radiative flux $\Fbot$ that is imposed at the bottom.
The red rolls correspond to the convective motions that develop in the layer 
middle.}
\label{fig:scheme}
\end{figure}

\begin{equation}
 K(T)=\Kmax\left[1+\Amp\dfrac{-\pi/2+\arctan(\sigma
T^+T^-)}{\pi/2+\arctan(\sigma e^2)}\right],
\label{eq:conductivity-profile1}
\end{equation}
with

\begin{equation}
 \Amp=\dfrac{\Kmax-\Kmin}{\Kmax},\quad T^{\pm}=T-\Tbump \pm e,
\label{eq:conductivity-profile2}
\end{equation}
where $\Tbump$ is the position of the hollow in temperature and
$\sigma$, $e$ and $\Amp$ denote its slope, width and relative amplitude,
respectively. We assume both radiative and hydrostatic equilibria, that is

\begin{equation}
 \left\lbrace
 \begin{aligned}
  \dfrac{dp_0}{dz} & = -\rho_0 g, \\
  \dfrac{dT_0}{dz} & = -\dfrac{\Fbot}{K_0(T_0)}, \\
 \end{aligned}
 \right.
 \label{eq:equil}
\end{equation}
where $\Fbot$ is the imposed bottom flux. Following paper I and II, we choose
the vertical dimension $L_z$ as the length scale, i.e. $[x]=L_z$, the top
density $\rhotop$ and top temperature $\Ttop$ as density and temperature scales,
respectively. The velocity scale is then $(c_p \Ttop)^{1/2}$ while time is
given in units of $[t] = L_z/ (c_p \Ttop)^{1/2}$.

As our aim is to investigate the convection-pulsation coupling, we must
ensure that convection will develop around the conductivity hollow. It
means that our model should satisfy to the well-known Schwarzschild's
criterion for the convective instability given by

\begin{equation}
 \left| \dfrac{dT}{dz}\right| > \left|
\dfrac{dT}{dz}\right|_{\text{adia}} \text{\quad with\quad} \left|
\dfrac{dT}{dz}\right|_{\text{adia}} = \dfrac{g}{c_p}.
\end{equation}
Using Eqs.~(\ref{eq:equil}), it defines the following minimum value for
the imposed bottom flux $\Fbot$

\begin{equation}
 \Fbot \ge \dfrac{g K_0(T_0)}{c_p}.
 \label{eq:schwarz}
\end{equation}
This criterion thus implies either a decrease in the gravity $g$ or an increase
in the imposed bottom flux $\Fbot$. However, we have shown in Paper I that one crucial
criterion needed to obtain unstable modes excited by $\kappa$-mechanism is that
the ratio of a local thermal timescale over the mode period is approximately
$\od{1}$, this is the relation (\ref{eq:psi}) already given in the
Introduction. Because of this timescale criterion, the most
favourable setups were found to require large values in the gravity $g$
to obtain a sufficient stratification (see Paper I for further details). As a 
consequence, the only
acceptable way to have a convective zone in our model --while keeping
the efficiency of $\kappa$-mechanism-- is not decreasing the
gravity $g$ but rather increasing the bottom flux such that
to satisfy Eq.~(\ref{eq:schwarz}).

Figure~\ref{fig:scheme} is a sketch of our convective model.
The blue curve denotes the radiative
conductivity profile from Eq.~(\ref{eq:conductivity-profile1}) that
fulfills Schwarzschild's condition (Eq.~\ref{eq:schwarz}, represented
by the dashed black line), leading to the appearance of a convective
zone shaped by the large red rolls where $K_0(T_0)\le c_p\Fbot/g$.
Nevertheless, the need to increase the bottom flux to get a convective
zone leads to a numerical difficulty that was not present in the pure
radiative case. Indeed, we can roughly estimate the typical velocity
$u_{\text{conv}}$ of a convective element by using a mixing-length
argument of the form

\begin{equation}
u_{\text{conv}} \sim \left( \frac{\Fbot}{\rho_0}\right)^{1/3}.
\end{equation}
With the typical values used in the former papers without convection,
that is $\rhotop \simeq 2.5 \times 10^{-3}$ and $\Fbot \simeq 2\times
10^{-2}$, one thus gets $u_{\text{conv}}\sim 2$ and therefore a Mach
number $\Mach \equiv u/c_s > 1$, with $c_s=\sqrt{\gamma R^* T}$ and
$T_{\text{top}}=1$. It means that the parameters used in these radiative
setups cannot be generalized to convective simulations as they
imply supersonic flows and shocks.

To avoid such intricate shocks in the simulation, one should reduce
the Mach number of the flow by both increasing the sound speed and
decreasing the typical convective velocity. Increasing $c_s$ is done
by simply doubling the value of the top temperature, that is, $\Ttop
=2$. As Schwarzschild's criterion (\ref{eq:schwarz}) imposes a large
value of the bottom flux $\Fbot$, we then decrease the convective velocity
$u_{\text{conv}}$ by increasing the top density to $\rhotop = 10^{-2}$,
then a factor 4 compared to the one in the radiative case.
We have also doubled the vertical extent of the layer to $L_z=2$ in
order to satisfy the timescale criterion (\ref{eq:psi}) while ensuring
that the convective zone is deep enough. Finally, concerning the
horizontal extent $L_x$ of the layer, the two following constraints
remain to be taken into account:

\begin{enumerate}
 \item If the aspect ratio $L_x/L_z$ is too small, the convective motions
 are too much constrained.
  Indeed, it is well known that a minimum
horizontal wavenumber $k_x$ is needed to trigger the convective instability
\cite[e.g.][]{Gough76}. As a consequence, dealing with small aspect
ratio $L_x/L_z$ can lead to a convective pattern too close to the
Rayleigh-B\'enard one, that is, some large periodic rolls along the
horizontal direction. In order to get a small-scale and as realistic
as possible convective pattern, one should therefore consider domains with 
large aspect ratios.
  \item Nevertheless, dealing with a large aspect ratio has an important
numerical cost. Indeed, with a sixth-order finite-difference code as
the one we are using, the mesh Reynolds number $\text{Re} = u \delta x /
\nu$ should not empirically be larger than $\sim 5$. It means that for
a given viscosity $\nu$ and spatial resolution, the horizontal extent
is anyway limited.
\end{enumerate}
Within these limits, the aspect ratio is kept constant in every
simulation discussed in this work, with $L_x / L_z = 4$ for $L_z=2$
(thus $L_x=8$).

Concerning the parameters of the conductivity profile, the central temperature
$\Tbump$ has been adjusted to satisfy the criterion given in
Eq.~(\ref{eq:psi}), that is $\Psi=1$ at the location of the hollow. As
in Papers I and II, its
relative amplitude $\Amp$ is kept constant with the same value,
that is $\Amp=0.7$. At last, the slope $\sigma$ and the width $e$ of
this profile are then set up to get as similar as possible profiles in the
different DNS computed in this study.

\begin{table*}
 \centering
 \begin{tabular}{cccccccccc}
 \toprule
 DNS & Gravity $\vec{g}$ & Flux $\Fbot$ & Conductivity $\Kmax$ & $\Tbump$ &
Width $e$ & Slope $\sigma$ &Viscosity $\nu$ & Frequency $\omega_{00}$ &
Rayleigh\\
 \midrule
 {\color{blue}\textbf{G8}} & 8 & $4.5\times 10^{-2}$ & $10^{-2}$ & 6
&1 & 1&$2.5\times 10^{-4}$ & $3.85$ & $1 0^5$\\
 G8V5 & 8 & $4.5\times 10^{-2}$ & $10^{-2}$ & 6 & 1 & 1 & $5\times
10^{-4}$ & $3.85$ & $5\times 10^4$\\
G8H95 & 8 & $4.5\times 10^{-2}$ & $9.5\times 10^{-3}$ & 6.5 & 1 & 1 & $5\times
10^{-4}$ & $3.84$ & $8\times 10^{4} $\\  
G8H9 & 8 & $4.5\times 10^{-2}$ & $9\times 10^{-3}$ & 7 & 1 & 1 & $5\times
10^{-4}$ & $3.83$ & $10^{5} $\\
{\color{OliveGreen}\textbf{G8H8}} & 8 & $4.5\times 10^{-2}$ & $8\times
10^{-3}$& 7.5 & 1 & 1.1 & $5\times 10^{-4}$ & $3.80$ & $2\times 10^{5}$ \\
G7  & 7 & $4\times 10^{-2}$ & $10^{-2}$ & 5.5 & 1.5 & 0.8 & $2.5\times 10^{-4}$
& $3.62$ & $8\times 10^{4} $ \\
G6   & 6 & $4\times 10^{-2}$ & $10^{-2}$ & 6 & 1.5 & 0.8 & $5\times 10^{-4}$ &
$3.35$ & $8\times 10^{4} $ \\
G6F7 & 6 & $3.7\times 10^{-2}$ & $10^{-2}$ & 5.7 & 1.5 & 0.8 & $3.5\times
10^{-4}$ & $3.36$ & $9\times 10^{4} $ \\
G6F7V4 & 6 & $3.7\times 10^{-2}$ & $10^{-2}$ & 5.7 & 1.5 & 0.8 & $4\times
10^{-4}$ & $3.36$ & $8\times 10^{4} $ \\ 
G6F5 & 6 & $3.5\times 10^{-2}$ & $10^{-2}$ & 5.5 & 1.5 & 0.8 & $3\times 10^{-4}$
& $3.36$ & $9\times 10^{4} $ \\
G6F5V4 & 6 & $3.5\times 10^{-2}$ & $10^{-2}$ & 5.5 & 1.5 & 0.8 & $4\times
10^{-4}$ & $3.36$ & $7\times 10^{4} $ \\ 
 
\bottomrule
 \end{tabular}
 \caption{Parameters of the numerical simulations done in this work. The two
colored bold-typed ones emphasised the two simulations discussed in the
following. For all these simulations, we assume $\Ttop=2$,
$\rhotop=10^{-2}$, $\Amp=0.7$ and $L_x/L_z=4$.}
 \label{tab:DNS}
\end{table*}

Table \ref{tab:DNS} summarizes the properties of the main numerical
simulations performed to study the convection-pulsation coupling.
The two coloured DNS, the G8 and G8H8 ones, are highlighted because we
will mainly focus on them in the following.
The penultimate column of this table contains the value of the
frequency $\omega_{00}$ of the fundamental unstable radial mode excited by
$\kappa$-mechanism, that lies between 3 and 4 for every DNS computed in this
study. The last column gives the value of the Rayleigh number that
quantifies the strength of the convective motions and it is given by

\begin{equation}
 \text{Ra} = \dfrac{g L_{\text{conv}}^4}{\nu \chi
c_p}\left|\dfrac{ds}{dz}\right|,
 \label{eq:rayleigh}
\end{equation}
where $L_{\text{conv}}$ is the width of the convective zone,
$\chi=K_0/\rho_0 c_p$ the radiative diffusivity and $s$ the entropy.
Tab.~\ref{tab:DNS} shows that
Rayleigh's numbers of the DNS presented in this paper are close
to $10^5$. This is well above the common critical values $\text{Ra}_c$ of this
number from which strong convection is known to develop, e.g. $\text{Ra}_c \sim
10^3$ for polytropic stratifications \citep{Gough76}.

\subsection{Dimensional values of the physical quantities}

The numerical simulations computed in this work are performed in dimensionless
units and can thus accommodate to a range of physical models. We can
nevertheless redimension our units to determine if they are close enough to the
realistic values of a typical Cepheid star. We assume that the top of our layer
is located under the surface at a temperature of 12000~K, it thus leads
to the temperature scale $[T]=6000~\text{K}$. For the G8 simulation given in
Tab.~\ref{tab:DNS}, we have approximately $T \in [2., 13.]$, meaning that our
layer covers a temperature range from 12000~K to 78000~K, that
well encompasses the second Helium ionisation zone. For a $5~\text{M}_{\odot}$
Cepheid star with parameters close to the ones of $\delta$-Cephei (i.e.
$\text{R}\sim 40~\text{R}_{\odot}$, $\text{L}\sim 2000~\text{L}_{\odot}$ and
$T_{\text{eff}}\simeq 6000~\text{K}$), such a temperature range represents
approximately $10\%$ of the star radius, leading to a length scale $[z] = 1.9
\times 10^{9}~\text{m}$\footnote{The values have been taken in a
$5~\text{M}_\odot$ stellar model provided by the Helas network on 
\url{http://www.astro.up.pt/helas/stars/}.}. The density
that corresponds to the temperature of
12000~K is approximately $4\times 10^{-5}~\text{kg}/\text{m}^3$, giving a
density scale $[\rho] = 4\times
10^{-3}~\text{kg}/\text{m}^3$ for the G8 simulation. In such a star, one has
$\rho_{\text{bot}}/\rho_{\text{top}} \simeq 170$, while our model has a
lightly smaller density contrast of about $\rho_{\text{bot}}/\rho_{\text{top}}
\simeq 140$.

The timescale that corresponds to the G8 simulation can then been computed with
the scaling $d/\sqrt{c_p \Ttop}$, leading to $[t] = 1.3\times
10^{5}~\text{s}$.
It means that a frequency $\omega_{00}=3.85$ for
the fundamental acoustic unstable mode corresponds in dimensional units
to a mode period of 2.5 days. This value is of the same order than the
period of $\delta$-Cephei, that is $5.36$ days.

With these scalings, we can also check if the gravity and fluxes of the DNS
are consistent with their stellar counterparts. The gravity $g=8$ in our units
then reads $g=0.84~\text{m}/\text{s}^2$. This value is very close to an estimate
of the surface gravity $g= GM/R^2$, that leads to $g=0.8~\text{m}/\text{s}^2$
for $\delta$-Cephei. Concerning the fluxes, $\Fbot = 4.5\times 10^{-2}$ then
becomes $5.2\times 10^{8}~\text{J}/\text{s}/\text{m}^2$. A good estimate of the
real flux is obtained with $F= L / 4\pi R^2$, leading for $\delta$-Cephei to
$7.7\times 10^{7}~\text{J}/\text{s}/\text{m}^2$. Our simple model thus
overestimates the heat flux but in a rather moderate way with respect to what is
usually done in global DNS \citep[e.g.][]{Bran00}. Such an accuracy in the
heat flux is in fact inherent in the study of the $\kappa$-mechanism as the
thermal timescale has to be close enough to the dynamical one to fulfill the
criterion (\ref{eq:psi}).

We can roughly estimate the kinematic viscosity of $\delta$-Cephei with
the following law $\nu \simeq 1.2\times10^{-17} T^{5/2} /
\rho~\text{m}^2/\text{s}$ \citep[see][]{Chapman54},
that gives $\nu = 2.5\times 10^{-3}~\text{m}^2/\text{s}$ in the layer we focus
on. A value $\nu_{\text{DNS}}=2.5\times 10^{-4}$ in the DNS
becomes $\nu_{\text{DNS}}=6\times 10^{9}~\text{m}^2/\text{s}$. We thus
overestimate the viscosity by a factor $10^{12}$. Such an enormous value is a
limitation of every DNS that cannot account for the small scale of
convection because of the limited numerical resolution
\citep[e.g.][]{Chan86}. As
a consequence, the viscosity is overestimated and leads to significantly lower
Rayleigh and Reynolds numbers than the real ones.

Table~\ref{tab:dimension} sums up the differences between the stellar
parameters of a Cepheid of $5~\text{M}_{\odot}$ with the dimensional units of
our local DNS. We notice that the values of our parameters are self-consistent
as we both recover temperature and density variations that are compatible with the
real ones, while keeping oscillation period, flux and gravity close enough to
their stellar counterparts.

\begin{table*}
\centering
\begin{tabular}{ccccc}
\toprule
Physical quantity & DNS dimensionless values & Scaling & DNS dimensional values
& Stellar values \\
\midrule
Length $L_z$ & 2 & $d$ & $1.9\times10^9$ m & $1.9\times 10^9$ m \\
$\Ttop$ & 2 & $\Ttop$ & 12000~K & 12000~K \\
$T_{\hbox{\scriptsize bot}}$ & 13 & $\Ttop$ & 78000~K & 78000~K \\
$\rhotop$ & $10^{-2}$ & $\rhotop$ & $4\times10^{-5}~\text{kg}/\text{m}^3$ &
$4\times10^{-5}~\text{kg}/\text{m}^3$
\\
$\rho_{\hbox{\scriptsize bot}}$ & 1.4 & $\rhotop$ &
$5.7\times10^{-3}~\text{kg}/\text{m}^3$
& $7\times10^{-3}~\text{kg}/\text{m}^3$ \\
Oscillation period & 1.63 & $d / \sqrt{c_p\Ttop}$ & $2.52$ days & $5.36$
 days \\
Gravity $g$ & 8 & $c_p\Ttop/d$ & $0.84~\text{m}/\text{s}^2$ &
$0.8~\text{m}/\text{s}^2$ \\
Flux & $4.5\times 10^{-2}$ & $\rhotop(c_p\Ttop)^{3/2} $ & $5.2\times 10^{8}
~\text{J}/\text{s}/\text{m}^2$ &
$7.7\times 10^{7}~\text{J}/\text{s}/\text{m}^2$ \\ 
Viscosity $\nu$ & $2.5\times 10^{-4}$ & $d\sqrt{c_p\Ttop}$ & $6\times
10^{9}~\text{m}^2/\text{s}$ & $2.5\times 10^{-3}~\text{m}^2/\text{s}$ \\
\bottomrule
\end{tabular}
\caption{Comparison between the main physical quantities given in
both the dimensionless units of the DNS and the corresponding dimensional
values, and their stellar counterparts for a $5~\text{M}_{\odot}$ Cepheid.}
\label{tab:dimension}
\end{table*}

\subsection{Numerical methods}

As in Papers I and II, the linear stability analysis of the initial
setup (\ref{eq:equil}) has been computed
thanks to the LSB code \citep[Linear Solver Builder,][]{LSB}. This
spectral solver, based on the iterative Arnoldi-Chebyshev algorithm
\citep{Arnoldi1951}, gives
both the complex eigenvalues $\lambda=\tau + i\omega$ and
corresponding eigenfunctions $\psi$ of a generalized eigenproblem of the form

\begin{equation}
A\vec{\psi}=\lambda B\vec{\psi},
\label{eq:eigenv}
\end{equation}
where the sparse matrices $A$ and $B$ result from the projection of the
differential operators on the Gauss-Lobatto grid. We especially focus on the
acoustic modes that are found to be linearly excited by the $\kappa$-mechanism,
that is those of which the real part of the eigenvalue is positive, $\tau > 0$.
Furthermore, this linear stability analysis has been computed \emph{without} the
consideration of the convective flux perturbations. This approximation,
called the \emph{frozen-in convection}, means that only the radiative
contribution of the total flux perturbation is allowed to change during the 
mode pulsation. We will see further that this approximation is well-suited in
our case as the computed eigenfunctions overlap quite well with the vertical
profiles extracted from the nonlinear simulation done with convection.

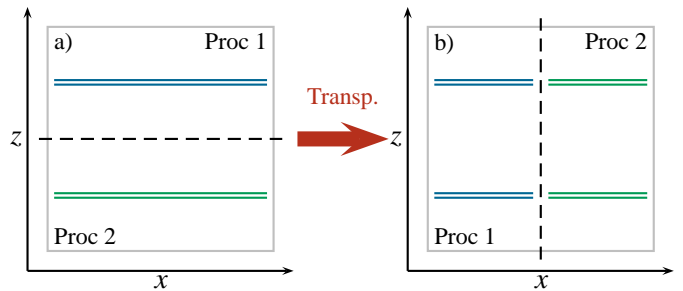
\begin{figure}
\centering
\begin{pspicture}(0,0)(9,3.7)
\psframe[linecolor=lightgray](0.5,0.5)(3.5,3.5)
\psline[linestyle=dashed](0.4,2)(3.6,2)
\rput[tl](0.6,3.4){a)}
\rput[tr](3.4,3.4){Proc 1}
\rput[bl](0.6,0.6){Proc 2}
\psline[linecolor=MidnightBlue](0.6, 2.78)(3.4, 2.78)
\psline[linecolor=MidnightBlue](0.6, 2.72)(3.4, 2.72)
\psline[linecolor=ForestGreen](0.6, 1.28)(3.4, 1.28)
\psline[linecolor=ForestGreen](0.6, 1.22)(3.4, 1.22)
\psframe[linecolor=lightgray](5.5,0.5)(8.5,3.5)
\psline[linestyle=dashed](7,0.4)(7,3.6)
\rput[tl](5.6,3.4){b)}
\rput[tr](8.4,3.4){Proc 2}
\rput[bl](5.6,0.6){Proc 1}
\psline[linecolor=MidnightBlue](5.6, 2.78)(6.9, 2.78)
\psline[linecolor=MidnightBlue](5.6, 2.72)(6.9, 2.72)
\psline[linecolor=MidnightBlue](5.6, 1.28)(6.9, 1.28)
\psline[linecolor=MidnightBlue](5.6, 1.22)(6.9, 1.22)
\psline[linecolor=ForestGreen](7.1, 2.78)(8.4, 2.78)
\psline[linecolor=ForestGreen](7.1, 2.72)(8.4, 2.72)
\psline[linecolor=ForestGreen](7.1, 1.28)(8.4, 1.28)
\psline[linecolor=ForestGreen](7.1, 1.22)(8.4, 1.22)

\psline[linewidth=0.2,linecolor=BrickRed]{->}(3.8,2)(5,2)
\rput[b](4.4,2.4){\color{BrickRed} Transp.}
\psline{->}(0.25,0.25)(3.75,0.25)
\rput[c](2,0.1){\large$x$}
\psline{->}(0.25,0.25)(0.25,3.75)
\rput[c](0.1,2){\large$z$}
\psline{->}(5.25,0.25)(8.75,0.25)
\rput[c](7,0.1){\large$x$}
\psline{->}(5.25,0.25)(5.25,3.75)
\rput[c](5.1,2){\large$z$}

\end{pspicture}
\caption{Sketch of the parallelisation of the ADI scheme. \textbf{a)} the
initial domain decomposition of the Pencil Code for a two-dimensional domain in
the $(x,\ z)$ plane. The coloured lines denote the data owned by the first
processor (blue lines) and by the second one (green lines). \textbf{b)} the
domain decomposition after the transposition needed to use the ADI solver.}
\label{fig:topology}
\end{figure}

Concerning the DNS, we have used again the Pencil Code\footnote{See
\url{http://www.nordita.org/software/pencil-code/} and
\cite{Pencil-Code}.}. This non-conservative code is an high-order
finite-difference code (sixth order in space and third order in time) that 
conserves the
physical quantities up to the discretization errors of the scheme. On
multiprocessor clusters, it takes advantage of
the MPI libraries (Message Passing Interface) that allow to distribute
the computing tasks on several processors having their own memory.
This code is fully explicit, except for the radiative diffusion term that is
solved implicitly thanks to an Alternate Direction Implicit
scheme (hereafter ADI) of our own already discussed in Paper I. However,
this new study has needed an improvement of this implicit solver as
the convective motions that develop in our simulations require its
parallelisation because of:

\begin{enumerate}
 \item the weakness of the mode growth rates $\tau \sim 10^{-3}$ computed
from the new setups that imposes to carry out the simulations
until a much longer time than in the purely radiative case.
 \item the convection itself that requires now a larger aspect ratio
ensuring an efficient development of all additional spatial scales. Larger
resolutions are thus compulsory to address the convection-pulsation
coupling.
\end{enumerate}

In its 2-D version, the Pencil Code has a domain decomposition along the
vertical direction such that the first direction of our ADI solver (the
horizontal one) well fits in this topology. On the contrary, the
second direction of the solver (the vertical one) is more tricky to 
parallelise as it 
needs a transposition of the data based on huge
communications between processors.
Figure~\ref{fig:topology} displays a sketch explaining
how the parallelisation of the ADI solver is done. Once the tridiagonal
system is solved in the second
direction, one transposes once
again the data to keep the original domain decomposition back.
With this powerful algorithm, the semi-implicit nonlinear simulations
of turbulent convection can then run on massively parallel clusters
with the large resolutions induced by high Rayleigh numbers (see 
Tab.~\ref{tab:DNS}). Typically, our 2-D
simulations of the convection-pulsation coupling were performed using
a mid resolution of about $512\times 512$ (i.e. 512 grid points in each
direction).

\subsection{\texorpdfstring{2-D DNS of the $\kappa$-mechanism with
convection}{2-D DNS of the k-mechanism with convection}}

\begin{figure*}
 \centering
 \includegraphics[width=12cm]{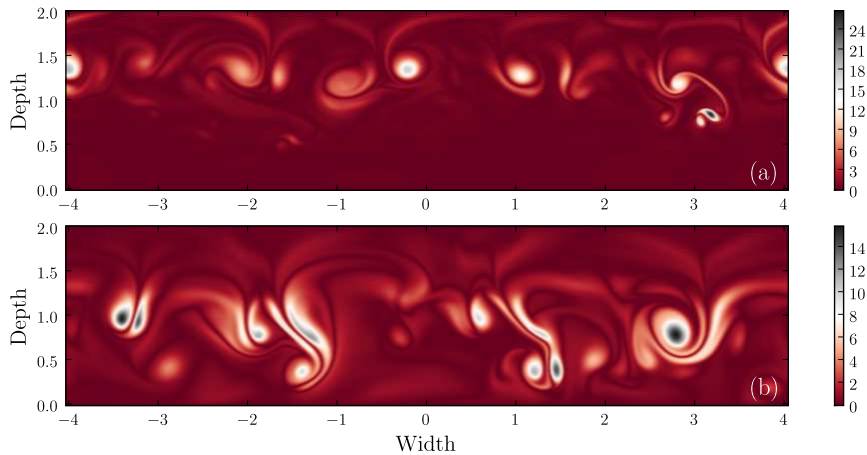}
 \caption{Snapshot of the modulus of the vorticity field $|\vec{\nabla}\times
\vec{u}|$ in the G8 (\textbf{a}) and in the G8H8 simulations
(\textbf{b}).}
 \label{fig:vort}
\end{figure*}

With the parallel ADI solver for the radiative diffusion implemented in the
Pencil Code (see Fig.~\ref{fig:topology}), we advance in time the following
hydrodynamic equations:
\begin{equation}
\left\lbrace
\begin{aligned}
\disp \frac{D \ln \rho}{Dt} = & -\Div \vec{u}, \\ \\
\disp \frac{D \vec{u}}{Dt} = & - \dfrac{1}{\rho}\na p
+ \vec{g} + 2\nu \lp \na \cdot \vec{\sf S} +\vec{\sf S}\cdot \na\ln\rho\rp, \\
\\
\disp \frac{DT}{Dt} = & \frac{1}{\rho c_v} \Div K\na T -(\gamma -1)T\Div\vec{u
} + 2\rho\nu \vec{\sf S}^2,
\end{aligned}
\right.
\label{eq:hydro}
\end{equation}
where $\rho,\ \vec{u}$ and $T$ denote the density, velocity and
temperature, respectively. The operator $D/Dt = \partial / \partial t +
\vec{u} \cdot \na$ is the usual total derivative, while $\vec{\sf S}$ is the
(traceless) rate-of-strain tensor given by

\begin{equation}
{\sf S}_{ij} = \frac{1}{2}\left(\frac{\partial u_i}{\partial x_j} +
\frac{\partial u_j}{\partial x_i}-\frac{2}{3} \delta_{ij} \Div \vec{u}\right).
\end{equation}
Finally, we impose that all fields are periodic in the horizontal
direction, while stress-free boundary conditions (i.e. $u_z=0$ and
$du_x/dz=0$) are assumed for the velocity in the vertical one. Concerning
the temperature, a perfect conductor at the bottom (i.e. flux imposed)
and a perfect insulator at the top (i.e. temperature imposed) are applied.

In order to ensure that both the nonlinear saturation and thermal
relaxation are well achieved,
the simulations were performed until very long times, typically
$t\gtrsim 3000$. With the eigenfrequencies computed from the
linear stability analysis, that is $\omega_{00} 
\simeq 3-4$ (see Tab.~\ref{tab:DNS}), this approximately corresponds to 1500
oscillation periods of the fundamental unstable acoustic mode. 

Figure~\ref{fig:vort} displays a snapshot of the modulus of
the vorticity field (i.e. $|\vec{\nabla}\times \vec{u}|$) at a given time 
for the two
simulations emphasised in Tab.~\ref{tab:DNS}, the G8
(Fig.~\ref{fig:vort}a) and
G8H8 ones (Fig.~\ref{fig:vort}b). The vorticity field highlights the convective
motions that are, as expected by the sketch in Fig.~\ref{fig:scheme},
superimposed to the radiative conductivity hollow
located approximately in the middle of the layer. Moreover, one
notes that the 
typical size of the convective plumes differs in the two simulations as the 
eddies have a
smaller scale in the G8 simulation than in the G8H8 one. Accordingly,
the overshooting of convective elements into the bottom stably stratified
layer is also more important in the G8H8 case than in the G8 one. It
suggests a lower value of the P\'eclet number in that case as a faster
thermalisation of the convective plumes at the interface between the
convective/radiative regions is associated with a larger penetration
extent \citep{Dintrans2009}.

\begin{figure}
 \centering
 \includegraphics[width=9cm]{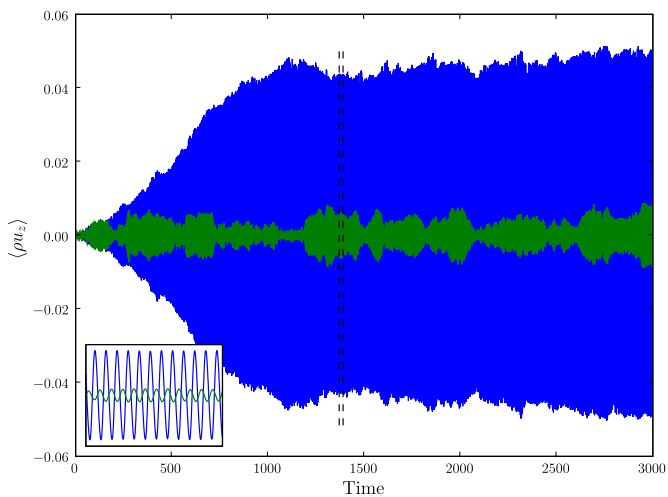}
 \caption{Temporal evolution of the mean vertical mass flux $\langle \rho u_z 
 \rangle$ for the two simulations G8 (solid blue line) and G8H8 (solid green
line). The two vertical dashed black lines define the boundaries of the zoom
displayed in the bottom left corner.}
 \label{fig:ts}
\end{figure}

Beyond the qualitative differences seen in the vorticity field, a good
way to compare the DNS is to study the temporal evolution of average
quantities, such as the vertical mass flux $\langle \rho u_z \rangle$,
where the brackets $\langle \cdots \rangle$ denote a global average.
Indeed, as we are doing simulations where both convective motions and
oscillations of acoustic modes are present, it is instructive to use a 
simple diagnostic
that roughly separates their relative contributions. As the 
convective
plumes have both ascending and descending movements, the average vertical
mass flux removes quite well their contribution
and one then gets a good tracer of the amplitude of
the acoustic modes {\it only}. Figure~\ref{fig:ts} shows 
the resulting temporal evolution of $\langle \rho u_z \rangle$
for the G8 simulation (blue line) and the G8H8 one (green line).
As displayed in the zoom in the bottom left corner, 
the time evolution shows an oscillating
behaviour in both cases due to the radial oscillations of the
fundamental (and unstable) acoustic mode excited by $\kappa$-mechanism. In the
G8 simulation, the amplitude experiences first an exponential growth until
reaching the nonlinear saturation regime for time $t\sim 1000$. One
notes that the transient duration is well compatible with the growth
rate of this mode that is about $\tau \sim 10^{-3}$. At first glance,
the temporal evolution of $\langle \rho u_z \rangle$ in the G8 simulation
is similar to what has been observed in the purely radiative simulations
of Paper II, that is, a linear growth of the amplitude before a saturation
on a finite value on a timescale $\propto 1/\tau$.

On the contrary, the dynamics of the G8H8 simulation radically differs
as the amplitude remains low compared to the radiative case and is highly
modulated in time. As a consequence, no clear nonlinear
saturation is observed in this case and this kind of behaviour has not
been found in the purely radiative models. It means that the acoustic
oscillations are more influenced by convective motions in the G8H8
simulation than in the G8 one. To study further the physics involved in this
coupling, we next perform a frequential analysis of these simulations.

\section{Frequential analysis}
\label{sec:freq}

\subsection{Fourier decomposition}

An efficient tool to determine the modes that are present in a numerical
simulation consists in taking first a double Fourier transform in
space and time of the vertical mass flux 

\[\rho u_z(x, z, t) \quad\Longrightarrow\quad \widehat{\rho u_z}(k_x, z,
\omega),\]
where $k_x=(2\pi/L_x)\ell$
denotes the horizontal wavenumber, with $\ell=[0,1,2,\dots]$, while
$\omega$ is the frequency. Second, one plots the power spectrum of
$\widehat{\rho u_z}$ in the plane $(z, \omega)$ for each value of
$\ell$ and ``shark fin profiles" emerge about definite frequencies
corresponding to eigenmodes \citep{Dintrans04}. The last operation
requires an integration over the vertical direction $z$ to get the final
mean spectra for each $\ell$ of the initial field $\rho u_z$.

\begin{figure}
 \centering
 \includegraphics[width=9cm]{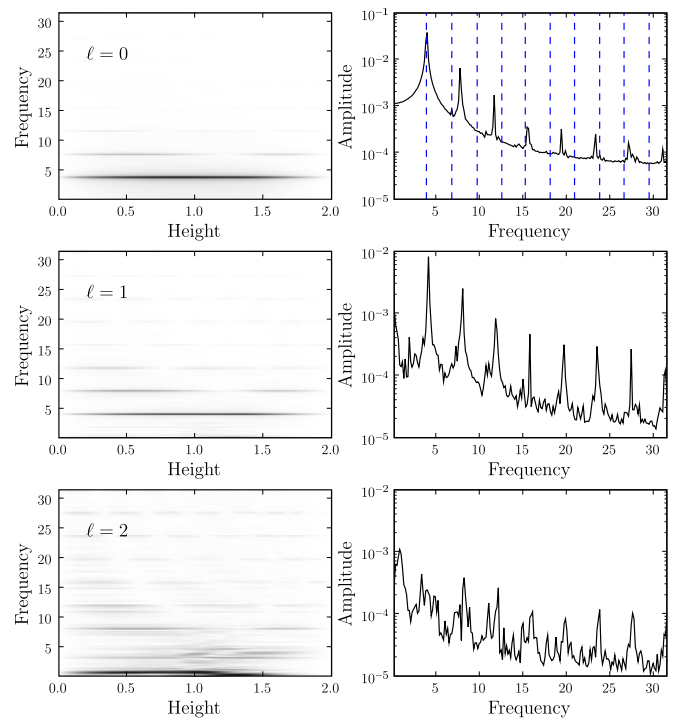}
 \caption{\textbf{Left :} temporal power spectrum of the vertical mass
flux $\widehat{\rho u_z}$ in the $(z,\ \omega)$ plane for the G8 simulation.
\textbf{Right :} the resulting spectrum after integrating over depth. In the
$\ell=0$ plane, the dashed blue vertical lines mark the position of the
frequencies of the overtones obtained with the linear stability analysis.}
 \label{fig:fourier_G8}
\end{figure}

\begin{figure}
 \centering
 \includegraphics[width=9cm]{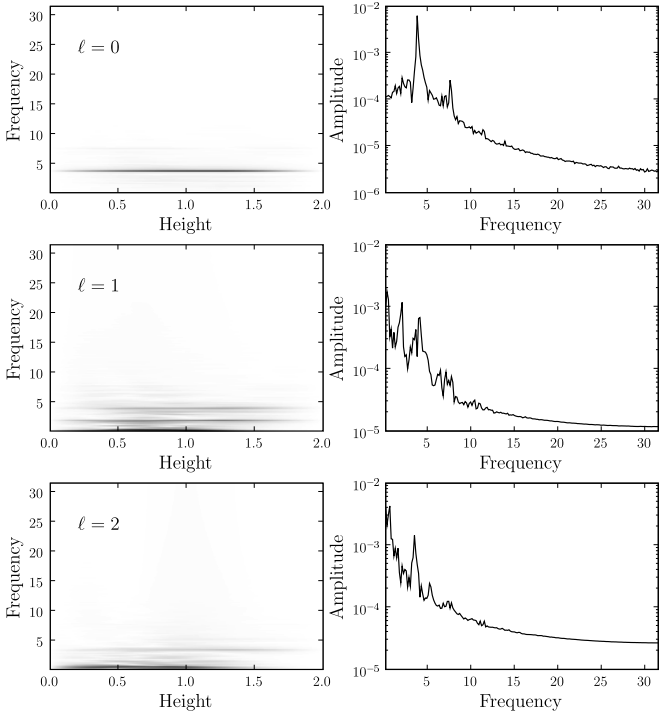}
 \caption{\textbf{Left :} temporal power spectrum for the vertical mass
flux $\widehat{\rho u_z}$ in the $(z,\ \omega)$ plane for the G8H8 simulation.
\textbf{Right :} the resulting spectrum after integrating over depth.}
 \label{fig:fourier_G8H8}
\end{figure}

Figure~\ref{fig:fourier_G8} displays the results of this Fourier analysis
applied to the G8 simulation, with the power spectrum of $\widehat{\rho
u_z}$ for the first three values of $\ell$ on the left side, and the
corresponding mean spectra on the right one. Many discrete peaks
about given frequencies are clearly noticed in the radial plane $\ell
= 0$ but also in the non-radial one $\ell = 1$, even the amplitudes
are weaker by an order of magnitude in this last case. As expected, the 
fundamental radial acoustic
mode $\omega_{00}=3.85$, which is the only one to be unstable linearly, has 
the largest amplitude in the $\ell =0$ plane. This signature is seconded by
numerous weaker amplitude peaks, of which the frequency is a multiple
of $\omega_{00}$. In fact, these peaks do no correspond to overtones
as they do not overlap with the theoretical eigenfrequencies sorted out
by the linear stability analysis (the computed overtones are displayed as
vertical dashed blue lines in Fig.~\ref{fig:fourier_G8}). They
rather correspond to harmonics of the fundamental mode and this is a
signature of the large amplitude of this mode. Indeed, the presence
of high-amplitude harmonics implies that the mode has a
sufficient amplitude for generating them through a nonlinear cascade. To
a certain extent, the same behaviour is obtained for the non-radial planes
$\ell=1$ and $\ell=2$, but with a much lower amplitude.

The temporal evolution of the mean vertical mass flux displayed in
Fig.~\ref{fig:ts}, which is similar to the ones obtained in the purely
radiative case, could indicate a nonlinear saturation of the G8
simulation based on 
a $2\text{:}1$ resonance between an unstable driving mode and a linearly stable
overtone that acts as a slave mode (see paper II for further details). But 
the Fourier analysis
in Fig.~\ref{fig:fourier_G8} emphasises that there is no overlap
between the frequencies of the DNS and the overtones ones. It
means that the coupling between different eigenmodes is not favoured in the 
G8 simulation
and a resonance-like phenomenon is not the physical process responsible
for its nonlinear saturation. Furthermore, the presence of numerous
large-amplitude harmonics is a guess of a mono-mode saturation,
such as the one observed for instance with Van der Pol's oscillators 
\citep[e.g.][]{vdp1,vdp2,Goupil84}.

The same Fourier analysis applied to the simulation G8H8 gives quite
different results compared to the G8 ones (Fig.~\ref{fig:fourier_G8H8}).
One first recovers the signature of some acoustic mode in the
$(z,\ \omega)$ plane, especially in the radial $\ell=0$ one
where the fundamental mode is well visible around the frequency
$\omega_{00}=3.80$. But the disappearance of the harmonics whatever the
value of $\ell$ means that the corresponding amplitudes of eigenmodes
are much lower than in the G8 case. And indeed, the amplitude of the
fundamental mode in the Fourier space is weaker by an order of magnitude
in the G8H8 simulation.

The differences in amplitude obtained previously in the time series in
Fig.~\ref{fig:ts} are thus confirmed by the results obtained in this
Fourier analysis: a high amplitude of the unstable acoustic mode,
observed in the G8 simulation, goes with numerous harmonics and
significant frequential signatures, while a weaker amplitude of this
mode, observed in the G8H8 simulation, is coupled with the disappearance of
these harmonics.

\subsection{Harmonic analysis}

\begin{figure}
 \centering
 \includegraphics[width=9cm]{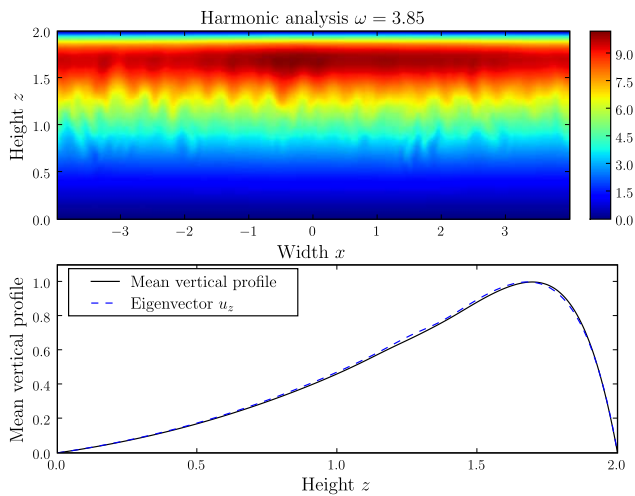}
 \caption{Harmonic analysis of the vertical
velocity field $u_z$ for the G8 simulation around the eigenfrequency 
$\omega_{00} = 3.85$. \textbf{Top:} amplitude of the modulus of the filtered 
velocity $|\hat{U}_{\omega_{00}}(x,z)|$. \textbf{Bottom:} normalised vertical
profile of the horizontal average of the filtered velocity
$|\hat{U}_{\omega_{00}}|$ (solid black line) and the corresponding velocity
eigenvector computed thanks to the linear stability analysis (dashed blue
line).}
 \label{fig:harmonic_G8}
\end{figure}

\begin{figure}
 \centering
 \includegraphics[width=9cm]{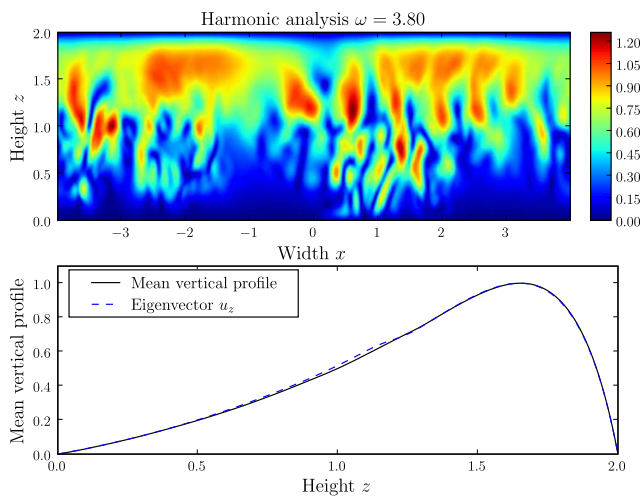}
 \caption{Harmonic analysis of the vertical
velocity field $u_z$ for the G8H8 simulation around the eigenfrequency 
$\omega_{00} = 3.80$. \textbf{Top:} amplitude of the modulus of the filtered 
velocity $|\hat{U}_{\omega_{00}}(x,z)|$. \textbf{Bottom:} normalised vertical
profile of the horizontal average of the filtered velocity
$|\hat{U}_{\omega_{00}}|$ (solid black line) and the corresponding eigenvector
of the linear stability analysis (dashed blue line).}
 \label{fig:harmonic_G8H8}
\end{figure}

In order to visualize the structure of an eigenmode oscillating in
a direct numerical simulation, hydrodynamicists have developed a diagnostic
tool called the ``harmonic analysis''. This method has been for instance
used to visualize internal wave attractors propagating in a non-separable
container \citep[e.g.][]{HBDM, grisouard08}. The idea is to filter out
the vertical velocity $u_z$ at a given frequency $\omega$ following
the relation


\begin{equation}
\hat{U}_{\omega}(x,z) =\dfrac{2}{T}\int_{t_1}^{t_2} u_z(x,z,t) e^{i \omega 
(t-t_1)} dt,
\end{equation}
where $t_2=t_1+nT$, with $n$ an integer, and $T = 2\pi/\omega$ is
the period. The obtained 2-D (complex) field $\hat{U}$ thus gives the
pattern that evolves in time in the simulation like $\cos \omega t$
or $\sin \omega t$. In \cite{grisouard08}, the chosen frequency was
imposed by a forcing term of the form $\cos \omega_{\text{forc}} t$
such that internal waves propagating in their simulation were necessarily
at this single frequency $\omega_{\text{forc}}$ (or its harmonics
$2\omega_{\text{forc}}, 3\omega_{\text{forc}}, \dots$).
In our case, the filtering frequency is of course given by the
acoustic modes excited by $\kappa$-mechanism and the resulting field
$\hat{U}_{\omega}(x,z)$ should correspond to the eigenvectors that are
solutions of the linear eigenvalue problem (\ref{eq:eigenv}).
We apply this method to the two simulations G8 and G8H8 for a filtering
frequency equal to the fundamental mode one, then $\omega=3.85$
and $\omega=3.80$ for the G8 and G8H8 simulation, respectively
(see Tab.~\ref{tab:DNS}). The resulting modulus $|\hat{U}_{\omega}(x,z)|$
is displayed in the figures~\ref{fig:harmonic_G8} (G8) and
\ref{fig:harmonic_G8H8} (G8H8).

For the G8 simulation, the
two-dimensional field $|\hat{U}_{\omega_{00}}|$ shows only variations in the
vertical
direction and has a quasi-periodic behaviour in the horizontal one
(upper panel in Fig.~\ref{fig:harmonic_G8}). This
well corresponds to the pattern that we are expecting for a radial
acoustic mode. Moreover, the convective plumes appear merely as very faint
hints under the form of several small wiggles. It means
that the
vertical motions at the frequency $\omega_{00}$ are dominated by the
acoustic radial mode. A definitive confirmation comes from the comparison
between the theoretical eigenvector and the horizontal average of
$|\hat{U}_{\omega_{00}}|$ given by

\begin{equation}
 |\hat{U}_{\omega_{00}}(z)| =\dfrac{1}{L_x}\int_{0}^{L_x} |
\hat{U}_{\omega_{00}}(x,z) |
dx.
\label{eq:aver}
\end{equation}
The result is shown in the bottom panel in
Fig.~\ref{fig:harmonic_G8} (solid black line). This vertical profile is 
then compared to the
eigenfunction $u_z$ computed in the linear stability analysis (dashed blue
line): the two lines overlap almost perfectly,
meaning that the motions at the frequency $\omega_{00}$ 
in the DNS have a vertical structure that corresponds to the unstable
fundamental mode.

The same harmonic analysis performed on the G8H8 simulation at the
frequency $\omega_{00}=3.80$ leads to different results
(Fig.~\ref{fig:harmonic_G8H8}). The 2-D modulus $|\hat{U}_{\omega_{00}}(x,z)|$ 
indeed differs significantly from the previous one as 
large horizontal variations are observed and
the expected invariance of the radial acoustic mode in that
direction is broken (upper panel in Fig.~\ref{fig:harmonic_G8H8}).
It means that convection
has now an important impact on the mode pattern
at this filtering frequency. The convective plumes strongly
affect the acoustic oscillations in this DNS and this is consistent with 
both the temporal
modulation of the vertical mass flux amplitude observed in
Fig.~\ref{fig:ts} and the lower Fourier amplitude of the fundamental
mode in Fig.~\ref{fig:fourier_G8H8}.
Despite this stronger coupling with convection, we nevertheless recover the 
structure of the unstable
acoustic mode after averaging $|\hat{U}_{\omega_{00}}|$ along the horizontal
direction
(lower panel in Fig.~\ref{fig:harmonic_G8H8}). The agreement between the
eigenvector and the vertical profile is rather remarkable but not really
surprising as the average in the horizontal direction exactly amounts
in the smoothing of the convective perturbations.

The frequential analysis developed in this section (Fourier
decomposition in the plane $(z,\ \omega)$ and the harmonic filtering)
confirms the discrepancies that were observed in the temporal
evolutions of the mean vertical mass flux $\langle \rho u_z \rangle$
(Fig.~\ref{fig:ts}): the acoustic oscillations are not influenced
by convection in the G8 simulation, whereas the
convective motions have a strong effect on the modes evolution in the
G8H8 simulation. We will now investigate in more details this effect by
the means of projections onto an acoustic subspace that give the temporal
evolution of the kinetic energy embedded in the pulsations.

\section{Energy of oscillations}
\label{sec:nrj}

\subsection{The projection method}

\begin{figure}
 \centering
 \includegraphics[width=9cm]{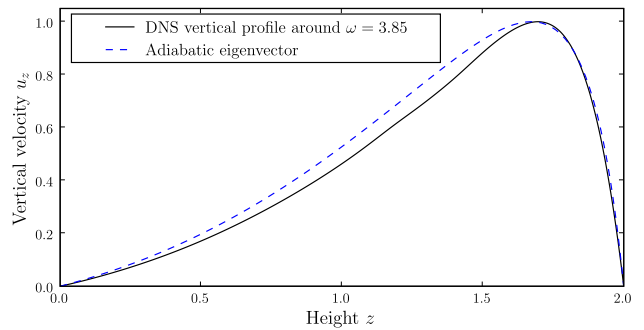}
 \caption{Comparison between normalised vertical velocity ($u_z$) profiles for
the fundamental mode ($n=0,\ \ell=0$) according to the harmonic analysis of the
DNS (solid black line) and the adiabatic eigenvector (dashed blue line).}
 \label{fig:adiabatic-eigenv}
\end{figure}

We have already used a projection method onto an acoustic subspace to
determine the energy contained in the acoustic modes of radiative
simulations (Paper II). This acoustic subspace was built from both normal and
adjoint eigenmodes that were solutions of the linear equations 
for the perturbations. The need to consider the adjoint problem, and not
only the regular one, was imposed by the non-hermiticity of the
oscillations operator. The radiative diffusion was indeed so large at the
surface of the equilibrium models that strong non-adiabatic effects made the 
regular eigenmodes non-orthogonal \citep[see also][]{Bogdan1993}. However, 
in the current 2-D simulations with convection, the setup
has significantly changed and the radiative diffusivity $\chi \propto
1/\rho$ is smaller than in the non-convective case (the top density has
been multiplied by a factor 4 in the convection setup, see
\S~\ref{sec:conv}).
The corresponding non-adiabatic effects at the surface are then weaker
and we use a simpler approach than the one in Paper II. This method,
based on the \textit{adiabatic} eigenvectors, has been validated in
\cite{Dintrans04} by measuring the wave field generated by the
oscillations of an entropy bubble in an isothermal atmosphere.

To use this approximation, we must first ensure that the vertical
velocity profiles in our 2-D DNS are sufficiently close to the adiabatic
eigenfunctions of the linear problem for the perturbations. This is
done in Fig.~\ref{fig:adiabatic-eigenv} where we compare the vertical
profile obtained in the previous harmonic analysis of the G8 simulation
(the solid black line in the lower panel in Fig.~\ref{fig:harmonic_G8}
and then also in Fig.~\ref{fig:adiabatic-eigenv}) to
the adiabatic eigenfunction $u_z$ (the dashed blue line in
Fig.~\ref{fig:adiabatic-eigenv}). The agreement between these
two profiles is rather comfortable, indicating that we can shape our
acoustic subspace from the adiabatic eigenvectors with a good
confidence.
This approach has the main advantage to simplify
the computation of the kinetic energy contents as \cite{lynden-bell}
have demonstrated that adiabatic eigenvectors
are mutually orthogonal for this particular scalar product

\begin{equation}
  \left\langle \vec{\psi_1}, \vec{\psi_2} \right\rangle =
  \displaystyle\int_z
\vec{\psi}^{\dag}_1\cdot\vec{\psi_2} \rho_0 dz.
\end{equation}
where the symbol $\dag$ denotes the Hermitian conjugate. The acoustic
subspace is then given by the following relation

\begin{equation}
\hat{\vec{u}}_{\ell}(z, t) = \displaystyle\sum_{n=0}^{+\infty}
c_{\ell n}(t)\vec{\psi}_{\ell n}(z)\quad \text{with}\quad
 \left\langle\vec{\psi}_{\ell' n'},\vec{\psi}_{\ell n}\right\rangle =
\delta_{\ell \ell'}\delta_{n n'}.
\label{eq:projcoeff}
\end{equation}
This defines the projection of the Fourier transform
$\hat{\vec{u}}_{\ell}(z, t)$ of the velocity onto the adiabatic
eigenvector $\vec{\psi}_{\ell n}(z)$ of degree $\ell$ and radial order
$n$. The complex and time-dependent coefficient $c_{\ell n}(t)$
entering in this projection is computed from

\begin{equation}
 c_{ \ell n}(t) = \left\langle \vec{\psi}_{n\ell}, \hat{\vec{u}}_{\ell}
\right\rangle = \displaystyle\int_0^{L_z}
\rho_0\vec{\psi}^{\dag}_{\ell n}\cdot\hat{\vec{u}}_\ell dz.
\end{equation}
As an example, when a given eigenmode $(\ell,\ n)$ is present in the
simulation, this projection method leads to a projection coefficient
that behaves as $c_{ \ell n}(t) \propto e^{i\omega_{\ell n} t}$, with
$\omega_{\ell n}$ the mode frequency.

\subsection{The kinetic energy content}

The Parseval equality written in the Fourier space reads

\begin{equation}
 \displaystyle\int_V \rho_0 \vec{u}^2(x,y,z) dV = \int_k \rho_0
|\widehat{\vec{u}_k}|^2 d^3 k,
\end{equation} 
where $\widehat{\vec{u}_k}$ denotes the Fourier transform in space of $\vec{u}$. In
our case, we only perform a Fourier transform of the velocity field in
the horizontal direction, therefore

\begin{equation}
E_{\text{kin}}^{\text{tot}} =  \displaystyle\int_0^{L_z} \rho_0 \vec{u}^2(x,z)
dx dz
= \int_0^{L_z} \sum_{\ell = 0}^{+\infty} \rho_0
\left|\hat{\vec{u}}_{\ell}(z)\right|^2
dz.
\label{eq:projnrj}
\end{equation}
with still $k_x = (2\pi/L_x) \ell$.  
With Eqs.~(\ref{eq:projcoeff}-\ref{eq:projnrj}), this leads to

\begin{equation}
 E_{\text{kin}}^{\text{tot}}(t) = \displaystyle\sum_{\ell =
0}^{+\infty}\sum_{n = 0}^{+\infty} |c_{\ell n}(t)|^2.
\label{eq:nrjmode}
\end{equation}
This last equation means that the kinetic energy content of a single
acoustic mode is simply equal to the square of its amplitude coefficient.
In the 2-D simulations with convection, we have

\begin{equation}
 E_{\text{kin}}^{\text{tot}}(t) = E_{\text{kin}}^{\text{modes}}(t)+
E_{\text{kin}}^{\text{conv}}(t),
\end{equation}
where $ E_{\text{kin}}^{\text{modes}}$ and $E_{\text{kin}}^{\text{conv}}$ are
the kinetic energy embedded in acoustic modes and convective motions,
respectively. In our setup, the fundamental radial mode $(\ell =0,\
n=0)$ is the only one that is excited by $\kappa$-mechanism therefore
the acoustic energy is contained in low-degree and low-order modes
(accordingly with the power spectra
in Fig.~\ref{fig:fourier_G8}). We can thus restrict the analysis to
these modes only in the summations entering in Eq.~(\ref{eq:nrjmode}), leading 
to

\begin{equation}
 E_{\text{kin}}^{\text{tot}}(t) = \disp\sum_{\ell = 0}^{3}\sum_{n = 0}^{5}
|c_{\ell n}(t)|^2+ E_{\text{kin}}^{\text{conv}}(t).
\end{equation}
The ratio of the kinetic energy contained in acoustic modes can then be
defined as

\begin{equation}
 R(t) = \dfrac{\disp\sum_{\ell = 0}^{3}\sum_{n = 0}^{5}
|c_{\ell n}(t)|^2}{E_{\text{kin}}^{\text{tot}}(t) }.
 \label{eq:acousratio}
\end{equation}

\begin{figure}
 \centering
 \includegraphics[width=9cm]{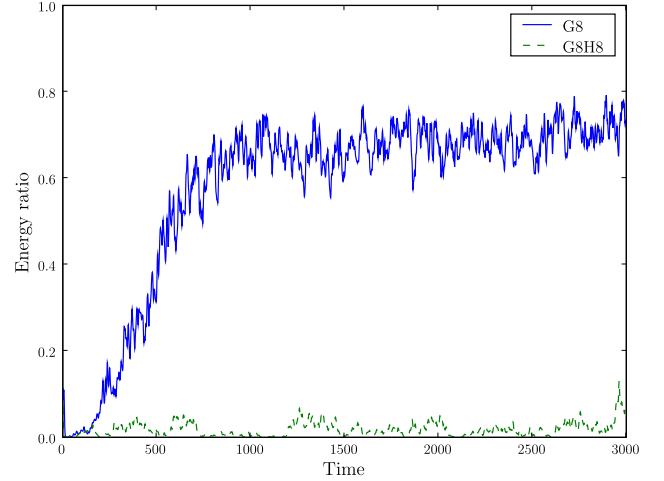}
 \caption{Temporal evolution of the acoustic kinetic energy ratio $R(t)$ for the
two simulations G8 (solid blue line) and G8H8 (dashed green line) according to
Eq.~(\ref{eq:acousratio}).}
 \label{fig:ratio}
\end{figure}

Figure~\ref{fig:ratio} displays the temporal evolution of this ratio
$R(t)$ for the two
simulations G8 (solid blue line) and G8H8 (dashed green line). 
In the G8 simulation, we see that the acoustic energy linearly increases until
its nonlinear saturation above time $t_{\text{sat}}\gtrsim 10^{3}$, in a similar way to 
what has been observed with the mean vertical mass flux in
Fig.~\ref{fig:ts}. This timescale $t_{\text{sat}}$
is still compatible with the linear
growth rate of the fundamental mode, that is, $\tau \sim 10^{-3}$ and
$t_{\text{sat}}\sim 1/\tau$.
Once this saturation is reached, the energy ratio remains large (i.e. $\gtrsim
70\%$, the remaining $30\%$ being in the convection) and this behaviour
is similar to the one observed in purely radiative
simulations (see Paper II for further details). In other words,
the acoustic oscillations are not much affected by the convective
motions in this simulation.

On the contrary, the acoustic energy ratio remains very
weak in the G8H8 simulation. Indeed, despite some transient increases
during which non-trifling values $R \simeq 10\%$ are obtained, the
average ratio is $R\lesssim
1-5\%$ and convective motions are responsible for the bulk of
the kinetic energy content. In this case, the radial oscillations excited by
$\kappa$-mechanism are thus quenched by convective plumes.
This situation is relevant to the red edge of the
classical instability strip, where the unstable acoustic modes are
supposed to be damped by the surface convective motions.

\subsection{Mean-field analysis}

Beyond the temporal evolution of the energy ratio displayed in
Fig.~\ref{fig:ratio}, it is interesting to precisely locate the zones where
kinetic energy is due to acoustic modes and convective motions, respectively. 
Towards this goal, a simple mean-field analysis is a valuable tool to
separate the mean-field component of the motion (the acoustic oscillations in
our setup) from its fluctuating part (here the convective plumes). Such approach
have been used for instance to check the mixing-length
theory of convection thanks to velocity correlations \citep{Chan1987}.

In our simulations, we are interested in the mean-field velocity correlations
that can account for the acoustic modes and the convective plumes. We first
separate the vertical velocity field into a mean part and a fluctuating one
\begin{equation}
 u_z = \langle u_z \rangle + u_z',
\end{equation}
where $\langle\cdots\rangle$ is an horizontal average. We then square this
expression and average along the horizontal direction to get
  
\begin{equation}
 \langle u_z^2 \rangle = \langle u_z \rangle^2 + \langle u_z'^2 \rangle.
  \label{eq:velocity-mean}
\end{equation}
This equation is the mean square velocity that can be split between an
acoustic contribution and a convective one, i.e. $\langle u_z^2 \rangle =
\text{acoustic} + \text{convection}$. To separate these different contributions,
we recall that $\langle u_z \rangle$ is a good proxy to measure
the acoustic oscillations, as the vertical convective motions
almost vanish after their averaging in the horizontal direction (ascending and descending plumes are comparable). The acoustic
contribution to Eq.~(\ref{eq:velocity-mean}) thus reads
 
\begin{equation}
\delta u^2_{\text{acous}}(z) =\langle u_z \rangle^2.
\label{eq:nrjacous}
\end{equation}
With Eq.~(\ref{eq:velocity-mean}), we then extract the contribution of the
convective motions to the mean square velocity field as

\begin{equation}
\delta u^2_{\text{conv}}(z) = \langle u_z'^2\rangle =\langle u_z^2 \rangle-
\langle u_z \rangle^2.
\label{eq:nrjconv}
\end{equation}
 
\begin{figure}
 \centering
 \includegraphics[width=9cm]{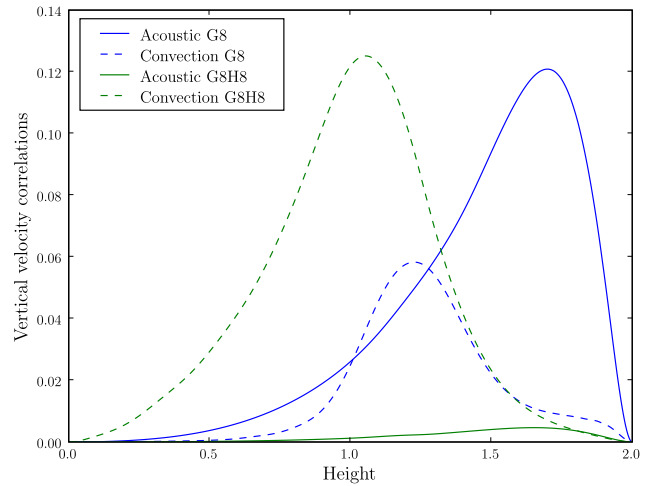}
 \caption{Vertical second-order velocity correlations for the simulations
G8 (blue lines) and G8H8 (green lines). The acoustic contribution
(Eq.~\ref{eq:nrjacous}) is displayed as solid
lines, while the convective one (Eq.~\ref{eq:nrjconv}) is plotted as
dashed lines.}
 \label{fig:vert_vel}
\end{figure}

Fig.~\ref{fig:vert_vel} displays the vertical profiles of $\delta
u^2_{\text{acous}}$ and $\delta u^2_{\text{conv}}$ for the simulations G8 (blue
lines) and G8H8 (green lines). In both simulations, one notes that the
maximum of the convective contribution (dashed lines) is approximately located
at the layer middle, corresponding to the radiative conductivity profile
displayed in Fig.~\ref{fig:scheme}, while the acoustic part (solid lines)
reaches its maximum close to the surface because of the eigenvector shape.

However, the acoustic contribution to the velocity
correlation for the G8 simulation is
larger than the convective one and a profile similar to the
velocity eigenfunction is obtained (see Fig.~\ref{fig:adiabatic-eigenv} for instance). On
the contrary, for the G8H8 simulation, this acoustic part $\delta
u^2_{\text{acous}}$ is much more weaker than the one due to convective motions. 

These trends can easily be linked to what was obtained previously with the 
projection formalism
(see Fig.~\ref{fig:ratio}). Indeed, the vertical profiles in
Fig.~\ref{fig:vert_vel} can be integrated to get a ratio between the
acoustic oscillations and the (total) mean square velocity $\langle u_z^2
\rangle$ as
  
\begin{equation}
 R_{\text{lim}} = \dfrac{\disp\int_0^{L_z} \delta u^2_{\text{acous}}\ dz}
{\disp\int_0^{L_z}
\langle u_z^2 \rangle\ dz},
\label{eq:ratio_prof}
\end{equation}
This last equation, derived from a mean-field analysis, contains the same 
physical
information than in Eq.~(\ref{eq:acousratio}) as the obtained ratio
is very close to the one computed with
the projection method in the saturated regime
(Fig.~\ref{fig:ratio}): $R_{\text{lim}} \simeq 75\%$ for the G8
simulation and $R_{\text{lim}} \simeq 3\%$ for the G8H8 one. The
second-order correlations $\delta
u^2_{\text{acous }}$ and $\delta u^2_{\text{conv}}$ are thus also a good
tool to separate
and locate the different contributions to the flow velocity.
  
The amount of energy contained in convective motions and in oscillations has
been studied thanks to both the projection formalism already used in purely
radiative simulations, and the second-order correlations of velocity. One
recovers the main physics underlined in the frequential analysis, that is, in
the G8 simulation the mode amplitude is strong and the oscillations clearly
dominate the motions in the flow, in opposition to the G8H8 simulation in which
convection is dominant.

\section{The mode quenching}
\label{sec:quench}

We are now going to focus on the physical process that lead to the different
behaviours obtained in our different DNS. In fact, as presented in
Tab.~\ref{tab:DNS}, the main physical parameters of these simulations are very
close: for instance, concerning the G8 and G8H8 simulation, there is mainly a
$20\%$ variation of the mean radiative conductivity $\Kmax$. Some of the 
parameters of the conductivity hollow (e.g. $\Tbump$ and $\sigma$) also differ 
as they have been adjusted to satisfy the criterion given in Eq.~(\ref{eq:psi}) 
in each DNS. We can classify the simulations given in Tab.~\ref{tab:DNS} in 
three different categories:
  
\begin{itemize}
 \item First, the simulations similar to the G8 one, in which the amplitude of
the unstable acoustic mode is important ($R\gtrsim 70\%$) and that doesn't seem
much affected by convection. This case concerns the simulation G7 of
Tab.~\ref{tab:DNS}.
 \item The second category, similar to G8H8, contains the DNS, in which the
amplitude of the acoustic mode is very weak ($R\lesssim 10\%$). In this case,
the oscillations are quenched by convective plumes that dominate the flow. This
case concerns also the G6 and G6F7 simulations of Tab.~\ref{tab:DNS}.
 \item At last, there are intermediate simulations, in which the amplitudes of
oscillations and convection are comparable ($R\simeq 30-50\%$). These
simulations also show important temporal modulation of the amplitude of
oscillations: i.e. the standard deviation of the ratio of energy is larger than
in previous categories. It mainly concerns the simulations G8H9 and G6F5 of
Tab.~\ref{tab:DNS}.
\end{itemize}

\subsection{The role of the convective flux}
 
\begin{figure}
 \centering
 \includegraphics[width=9cm]{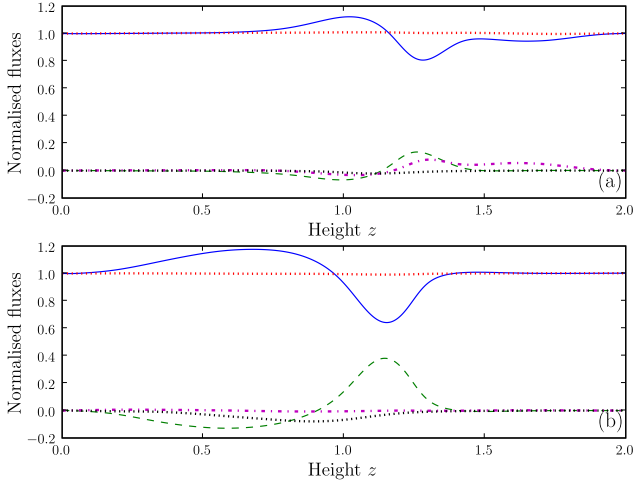}
 \caption{Mean vertical profiles of radiative $\Fr$
 (solid blue line), turbulent enthalpy $\Fc$ (dashed green line),
kinetic
$\Ft$ (dotted black line), modes $\Fs$ (dot-dashed magenta line) and
total ${\cal F}_{\text{tot}}$ (dotted red line) fluxes for the G8 \textbf{(a)}
and G8H8 \textbf{(b)} simulations. All fluxes have been normalised with
respect to the bottom value $\Fbot$.}
 \label{fig:fluxes}
\end{figure}

A first attempt in explaining these different physical behaviours
is to compare the heat transport between the various DNS that are
summarised in Tab.~\ref{tab:DNS}. Especially, the study of the respective amount of 
heat carried out by convection and radiation may lead to a natural
explanation to the mode quenching observed in the G8H8 simulation as,
from an intuitive point of view,
the higher the convective flux the stronger the quenching.

The vertical heat transport is ensured by the radiative $\Fr$, enthalpy $\Fh$ 
and kinetic $\Ft$  fluxes given by
\citep[e.g.][]{Hurlburt84}

\begin{equation}
\left\lbrace
\begin{aligned}
\Fr(z,t) &=  -\left\langle K(T) \nabla T \right\rangle, \\ 
\Fh(z,t) &=  c_p \left\langle \rho u_z T \right\rangle, \\ 
\Ft(z,t) &=  \dfrac{1}{2} \left\langle (u_x^2+u_z^2) \rho u_z \right\rangle,
\end{aligned}\right.
\label{eq:fluxes}
\end{equation}
where the brackets denote an horizontal average. The
enthalpy flux $\Fh$ can be developed in mean and fluctuating components
from the following decomposition for
the total temperature $T$

\begin{equation}
\begin{aligned}
 T(x,z,t) & = \langle T \rangle(z,t)+ T'(x,z,t), \\
          & = T_{\text{HS}}(z)+\theta(z,t)+ T'(x,z,t),
\end{aligned}
\end{equation}
where $T_{\text{HS}}$ is the hydrostatic background temperature profile,
$\theta$ is the temperature eigenfunction of the unstable acoustic mode and $T'$
is the turbulent fluctuating temperature around the horizontal average $\langle
T \rangle$. The enthalpy flux then reads

\begin{equation}
 \Fh(z,t) = c_p \left\langle \rho u_z T' \right\rangle + c_p\left\langle \rho
u_z\right\rangle\left(T_{\text{HS}}+\theta\right).
\end{equation}
We now average in time this last expression over a multiple of the mode period
to get

\begin{equation}
 \Fh(z) = \overline{\Fh(z,t)} = c_p \overline{\left\langle \rho u_z
T'\right\rangle}+  c_p \overline{\left\langle \rho
u_z\right\rangle\theta}+c_p
T_{\text{HS}}\overline{\left\langle \rho
u_z\right\rangle}.
\end{equation}
Finally, 
as there is no average mass flux over an oscillation period (i.e.
$\overline{\left\langle \rho u_z\right\rangle}=0$), the enthalpy flux
carried by both the convection and acoustic oscillations is given by

\begin{equation}
 \Fh(z)= c_p \overline{\left\langle \rho u_z T'
\right\rangle}+c_p\overline{\left\langle \rho
u_z\right\rangle\theta}.
\label{eq:enthalpy}
\end{equation}
The first term of the right hand side denotes the turbulent transport of
enthalpy usually defined as the convective flux \citep[e.g.][]{Hurlburt84},
while the second one corresponds to enthalpy transported by the
acoustic modes. 
This second contribution is usually negligible in simulations of
compressible convection because of the weakness of modes that are
driven by turbulent fluctuations in pressure
\citep{Bogdan1993}. In our case, the modes can efficiently be excited
through the $\kappa$-mechanism and their contribution to the
heat transport should be taken into account.
In order to determine
the net contribution of the acoustic modes to the total flux, we thus
separate the usual turbulent enthalpy part due to the convective plumes
(hereafter denoted $\Fc$) from the enthalpy transport due to
modes (hereafter $\Fs$). By taking the average in time of Eqs.~
(\ref{eq:fluxes}) and using Eq.~(\ref{eq:enthalpy}), one finally gets

\begin{equation}
\left\lbrace
\begin{aligned}
\Fr(z) & = -\overline{\left\langle K(T) \nabla T \right\rangle}, \\ 
\Fc(z) & = c_p \overline{\left\langle \rho u_z T' \right\rangle}, \\ 
\Ft(z) & = \dfrac{1}{2}\overline{\left\langle (u_x^2+u_z^2) \rho u_z
\right\rangle}, \\
\Fs(z) & = c_p\overline{\left\langle \rho u_z\right\rangle\theta}.
\end{aligned}\right.
\label{eq:fluxes_temp}
\end{equation}
The vertical profiles of these fluxes (normalised to the imposed
bottom value $\Fbot$) are given in Fig.~\ref{fig:fluxes} for the G8
(upper panel) and G8H8 (lower panel) simulations. For both DNS, the
bulk of the total flux is transported by the radiative flux, except in the
convective zone where $\Fc$ reaches $15\%$ of the total flux for the G8
simulation and $40\%$ for the G8H8 one. We also notice a larger overshooting in
the second DNS, as the downdrafts penetrating in the radiative zone are
stronger (see Fig.~\ref{fig:vort}). As expected, this significant
penetration is associated with both a negative convective flux and a
small value of the kinetic flux ($\sim 5\%$). The convective
fluxes in our DNS are nevertheless relatively weak compared to what is expected
to occur in Cepheid stars \citep[see e.g.][]{YKB98}. These limited values are a
direct issue of the rather moderate Rayleigh number numerically affordable in
such kind of DNS. 

Concerning $\Fs$, one notes that it is hardly as
important as the convective flux in the G8 simulation, while it is almost
vanishing in the G8H8 one. In the G8 simulation where the amplitude of the
unstable acoustic mode is the largest, about $10\%$ of the heat flux is carried
out by acoustic modes. This quantity is thus another good tracer of the
significance of the amplitude of the acoustic modes compared to the convective
motions, as $\Fs$ becomes significant when the amplitude of the acoustic
oscillations is strong enough.

\begin{figure}
 \centering
 \includegraphics[width=9cm]{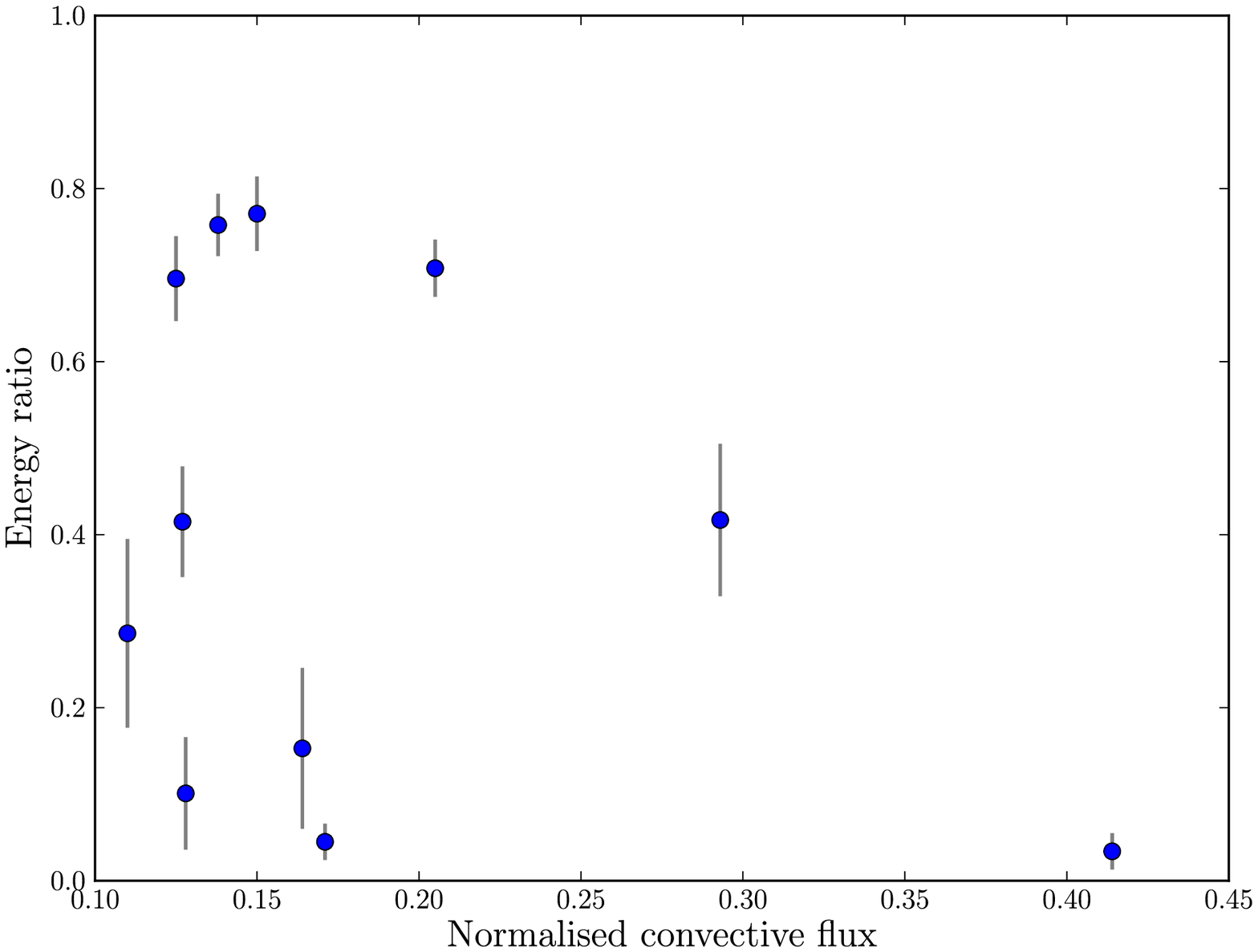}
 \caption{Ratio of the kinetic energy in acoustic modes displayed as a
function of the maximum of the normalised convective flux $\Fc$ for the
different DNS presented in Table~\ref{tab:DNS}. The blue dots correspond to the
mean kinetic ratio obtained with the projection method, while the vertical gray
bars represent the standard deviation.}
 \label{fig:fconv}
\end{figure}

The differences in the amount of heat transported by turbulent convection may be
a first hint on the physical origin of the mode quenching observed in our DNS.
To confirm this hypothesis, we have thus gathered the values of the maximum of
the normalised convective flux for all the DNS in Tab.~\ref{tab:DNS}.
Figure~\ref{fig:fconv} displays the amount of energy contained in acoustic
modes with respect to these maximum values.
For moderate to stronger convective fluxes, we recover a
significant relation between the fraction of $\Fc$ and the efficiency of the
$\kappa$-mechanism, that is, the energy ratios in acoustic modes become smaller
with increasing convective fluxes. But this correlation disappears in the
region of smaller convective fluxes with $\Fc\lesssim 20\%$. Indeed,
one observes here that the efficiency of the $\kappa$-mechanism is 
independent of the value of the convective flux as the modes may be
either quenched or excited for similar values of $\Fc$.

Fig.~\ref{fig:fconv} thus emphasises that the physical origin of the mode
quenching does not solely rely on the amount of the convective flux, as
the acoustic oscillations are also damped in DNS where the convective 
transport is inefficient.
This result indicates that other underlying physical processes
are involved in this mode quenching.

\subsection{The role of the stratification}

\begin{figure}
 \centering
 \includegraphics[width=9cm]{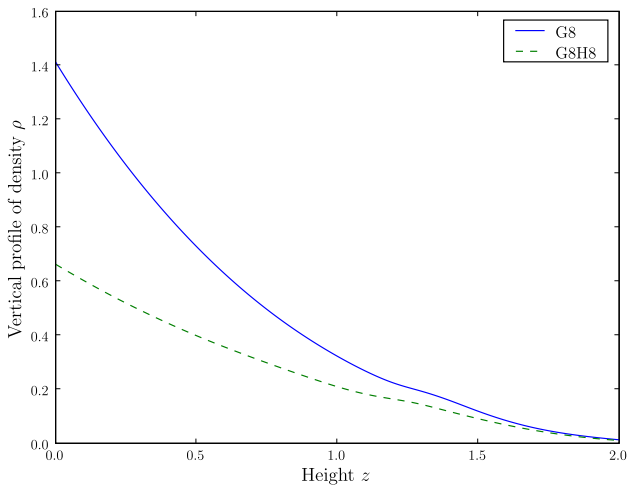}
 \caption{Mean vertical profiles of density $\rho$ for the
G8 simulation (solid blue line) and the G8H8 simulation (dashed green line).}
 \label{fig:rhoprof}
\end{figure}

The differences between the physical parameters of the DNS given in
Table~\ref{tab:DNS} seem at a first look negligible. However the variations of 
gravity $g$, flux
$\Fbot$ and radiative conductivity $\Kmax$ lead to significantly modified
equilibrium fields, especially concerning the density profile. The mean
vertical profile of density is close to the hydrostatic equilibrium, that reads

\begin{equation}
 \dfrac{d\ln\rho}{dz} =
\dfrac{1}{T}\left(\dfrac{\Fbot}{K(T)}-\dfrac{g}{R^*}\right),
\label{eq:hydrostatic_rho}
\end{equation}
where the radiative conductivity $K(T)$ is given by
Eq.~(\ref{eq:conductivity-profile1}). Increasing $\beta=\Fbot/\Kmax$, or
decreasing the gravity $g$ then leads to a weaker derivative of the density.
Fig.~\ref{fig:rhoprof} displays the mean vertical density profile for the
simulations G8 (solid blue line) and G8H8 (dashed green line). It shows that a
$20\%$ change of the radiative conductivity leads to significantly different
density values, especially at the bottom of the layer where one notes 
a factor of two between the simulations.

To compare the differences in the density stratification between the
simulations of Tab.~\ref{tab:DNS}, we may use an analytical approximation of
the density scale $H_{\rho} \equiv -( d\ln\rho/dz)^{-1}$ at the layer middle.
Starting from Eq.~(\ref{eq:hydrostatic_rho}), one can approximate the
temperature profile by

\begin{equation}
 \dfrac{dT}{dz} = -\dfrac{\Fbot}{K(T)} \quad\Longrightarrow\quad
T(z) \simeq
-\beta(z-L_z) + \Ttop,
\end{equation}
where we suppose that the mean temperature profile is linear, that is the
conductivity profile is neglected and assumed to be $K(T) \simeq \Kmax$. The
hydrostatic equilibrium then reads

\begin{equation}
 \dfrac{d\ln\rho}{dz} \simeq \dfrac{1}{-\beta(z-L_z) + \Ttop}\left(\beta -
\dfrac{g}{R^*}\right).
\end{equation}
We can then approximate the density scale $H_{\rho}$ at the layer middle by

\begin{equation}
 {H_{\rho}}_{\text{middle}} \simeq -\dfrac{\beta L_z/2+\Ttop}{\beta-g/R^*}.
 \label{eq:hrho}
\end{equation}
This approximation allows us to compare the different simulations of
Tab.~\ref{tab:DNS}. The results given in Tab.~\ref{tab:hrho} emphasises a 
correlation between the density profile and the ratio of energy contained in
acoustic modes, that is the weaker $H_{\rho}$ the larger $R_{\text{lim}}$. On
the contrary acoustic modes are quenched by convective motions if the value of
$H_{\rho}$ increases (simulations G8H8 or G6 for instance).

\begin{table}
 \centering
 \begin{tabular}{ccc}
 \toprule
 Simulation & ${H_{\rho}}_{\text{middle}}$ & $R_{\text{lim}}$ \\
 \midrule
 {\color{blue}\textbf{G8}} & 0.42 & $76\%$\\
  G8V5 & 0.42 & $77\%$ \\
  G8H95 & 0.44 & $71\%$ \\
 G8H9 & 0.47 & $42\%$ \\
 {\color{OliveGreen}\textbf{G8H8}} & 0.53 & $3\%$ \\
 G7   & 0.44 & $70\%$ \\
 G6   & 0.55 & $4\%$ \\
 G6F7 & 0.50 & $15\%$ \\
 G6F7V4 & 0.50 & $10\%$ \\
 G6F5 & 0.48 & $41\%$ \\
 G6F5V4 & 0.48 & $28\%$ \\
  \bottomrule
 \end{tabular}
 \caption{Values of the density scale at the layer middle according to
Eq.~(\ref{eq:hrho}) and ratio of energy contained in acoustic modes according
to Eq.~(\ref{eq:acousratio}).}
 \label{tab:hrho}
\end{table}

\begin{figure}
 \centering
 \includegraphics[width=9cm]{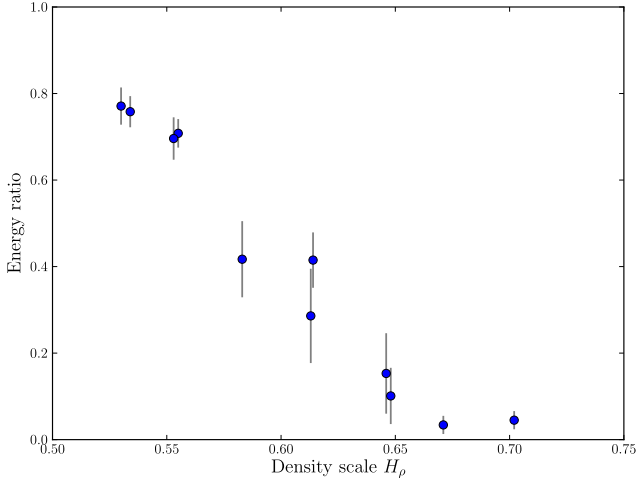}
 \caption{Ratio of the kinetic energy in acoustic modes displayed as a function
of the mean density scale $\langle H_{\rho} \rangle$ for the different DNS
presented in Table~\ref{tab:DNS}. The blue dots correspond to the mean kinetic
ratio obtained with the projection method, while the vertical gray bars
represent the standard deviation.}
 \label{fig:hrho}
\end{figure}

We can also extract the real values of $H_{\rho}$ directly from the DNS field
in order to check that the results obtained in Tab.~\ref{tab:hrho} are
confirmed without the constant radiative conductivity profile approximation
done in Eq.~(\ref{eq:hrho}). Fig.~\ref{fig:hrho} displays the ratio
$R(t)$ as a function of the mean density scale $\langle H_{\rho}
\rangle$ for the different DNS. Blue dots corresponds to $\overline{R(t)}$
(temporal average) and vertical bars to the standard deviation of $R(t)$. We
recover the correlation between the value of the density scale and the
significance of
pulsations compared to the convective motions. What's more, Fig.~\ref{fig:hrho}
emphasises the three different behaviours presented before, as one gets the two
extrema (high amplitude of modes or mode-quenching by convection) but also
``mixed''-simulations with $R\simeq 30-50\%$ and very large standard
deviations meaning that an important modulation of pulsations occurs. 

\subsection{Spectrum and integral scale}

Fig~\ref{fig:hrho} seems to indicate that the density stratification plays an
important role on the stabilisation of acoustic modes excited by
$\kappa$-mechanism. According to the mixing length theory \citep{BV1,BV2}, more
important values of $H_{\rho}$ means that the typical size of convective eddies
increases. This assumption is besides displayed on Fig.~\ref{fig:vort}, where
the differences between the size of the vortices in the G8 and G8H8 simulations
are noticeable. To determine the role played by the different scales of the
flow onto the pulsations, we now focus on the power spectra of the different
DNS.

\begin{figure}
 \centering
 \includegraphics[width=9cm]{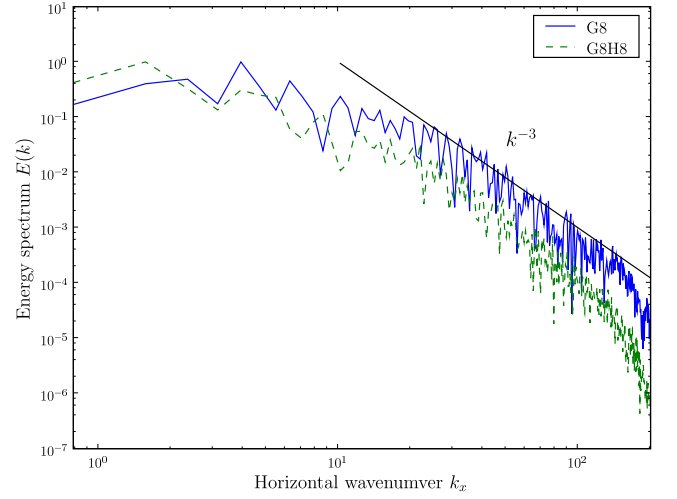}
 \caption{Normalised kinetic energy power spectrum in the middle of the
convective zone for the G8 simulation (solid blue line) and the G8H8 simulation
(dashed green line). The kinetic energy spectrum $E(k) \sim k^{-3}$ has been
superimposed as a solid black line.}
 \label{fig:spectrum}
\end{figure}

The power spectrum is obtained with a Fourier transform of the kinetic energy
along the horizontal direction :

\begin{equation}
 E(k,z) = \int_0^{L_x} \rho (u_x^2+u_z^2) e^{-i k x} dx.
 \label{eq:spectrum}
\end{equation}
The normalised power spectrum at the middle of the convective layer ($z\simeq 
1.2$) is then displayed on Fig.~\ref{fig:spectrum} for the two simulations G8
and G8H8. One notes that the solid blue line, corresponding to G8, reaches its
maximum to an higher value of $k_x$ than the dashed green line (G8H8). The
energy spectrum is thus shifted to smaller scale in the G8 simulation than in
the G8H8, in which the kinetic energy is stored in larger scales. It
corresponds to what is observed on the vorticity fields of Fig.~\ref{fig:vort},
where the convective eddies are larger in the G8H8 simulation than in the G8
one.

We can also mention that we recover in the kinetic energy spectra a scaling law
$E(k) \sim k^{-3}$. This result corresponds to the standard dimensional
analysis of the enstrophy cascade in two-dimensional turbulence
\citep[e.g.][]{Lesieur}.

From Eq.~(\ref{eq:spectrum}), we can determine the integral scale,
that is the scale associated with the most energetic structures of the flow.

\begin{equation}
  \ell_{\text{int}}(z) = \dfrac{\displaystyle\int_0^\infty E(k,z) k^{-1}
dk}{\displaystyle\int_0^\infty E(k,z) dk}.
 \label{eq:lint}
\end{equation}

\begin{figure}
 \centering
 \includegraphics[width=9cm]{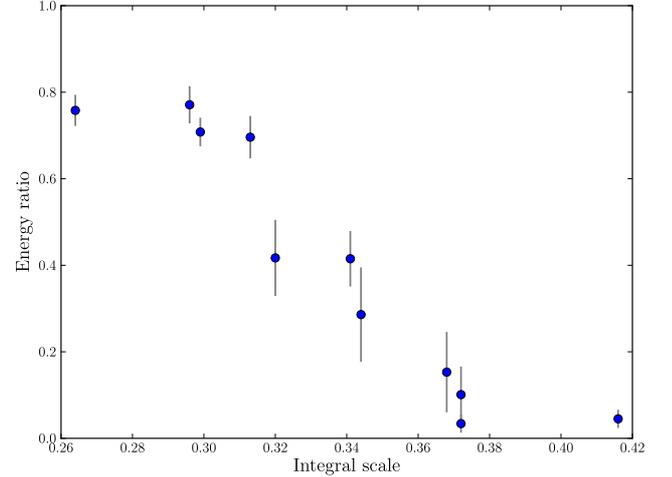}
 \caption{Ratio of the kinetic energy in acoustic modes displayed as a function
of the integral scale (Eq.~\ref{eq:lint}) for the different DNS presented in
Table~\ref{tab:DNS}. The blue dots correspond to the mean kinetic ratio obtained
with the projection method, while the vertical gray bars represent the standard
deviation.}
 \label{fig:integral}
\end{figure}

Once again, we compute the integral scale at the middle of the convective zone
for the different DNS of Tab.~\ref{tab:DNS}. Fig.~\ref{fig:integral} displays
the value of $\overline{R(t)}$ with respect to the integral scale
$\ell_{\text{int}}$. The vertical gray bars emphasize the standard deviation of
$R(t)$. Fig.~\ref{fig:integral} illustrates the same trend as in
Fig.~\ref{fig:hrho}: the larger the integral scale, the smaller the energy
ratio. The integral scale is in fact another way to get the information
obtained with the $H_{\rho}$ length scale, but relies on the properties
of the turbulence of the flow.
 
 The comparison of these length scales in our DNS is a first hint on the physics
responsible for the mode quenching. In the light of
Figs.~\ref{fig:hrho}-\ref{fig:integral}, the density stratification seems to
have an important impact on how the oscillations develop in the presence of
convection. With the increase of the size of the eddies (increase of
$\ell_{\text{int}}$), the acoustic part in the energy budget is gradually
vanishing. 

We may explain this mode quenching by the spatial distribution of the
convective plumes: large eddies entail an important ``screening effect'' onto
the acoustic radial mode. One can imagine that the amplitude of such mode is
affected by the motions that are not coherent to its vertical structure: in
this case large eddies with significant amount of energy distributed both in
radial and non-radial motions is incoherent with the ``main'' velocity field due
to the radial purely pulsations. In the less stratified simulations, such as
G8H8, the convective plumes are large enough and present an
important horizontal filling-factor that leads to a velocity field incompatible
with the acoustic mode structure.

Such screening effect, or such spatial distribution of the convective plumes
, have been already found to be responsible of the modulation and quenching of
gravity modes in DNS of the penetrative convection \citep{dintrans05}. In our
DNS, where oscillations are sustained by a physical process acting
continuously, the amplitude of pulsations (and their nonlinear saturation) may
thus be ruled by the extent of the screening of large vortices onto the purely
radial velocity field.

\section{Conclusion}
\label{sec:conclu}

In this paper, we have investigated the convection-pulsation coupling thanks
to nonlinear two-dimensional direct numerical simulations (DNS). Despite their 
intrinsic limitations \citep[weak pressure contrasts
across the computational domain or the need of large viscosities leading to 
unrealistic values of Rayleigh's number, see e.g.][]{Bran00}, DNS are without a doubt
an important way to address this interaction as they fully account for the
nonlinearities.

This study follows our former work on the $\kappa$-mechanism in
purely radiative simulations, where we studied both the instability and its
nonlinear saturation \citep[][Papers I and II, respectively]{paperI, paperII}.
We have then again modelled the ionisation region
responsible for the oscillations of classical Cepheids by a hollow in
the radiative
conductivity profile and \emph{local} simulations of a
perfect gas layer centered around this ionisation bump have been
performed.
The initial physical setup has been however slightly modified to get a 
convective zone
superimposed with the hollow in radiative conductivity.
Furthermore, some important numerical improvements have been developed
(mostly the parallelisation of the ADI solver) as the taking into account
of strong convective motions coupled with acoustic oscillations requires
larger spatial resolutions than the ones used in the purely radiative case.

Using this parallel solver, we have performed 2-D DNS of the
convection-pulsation coupling in which the oscillations are sustained by a
continuous physical process based on the $\kappa$-mechanism. Convective motions that
develop in these simulations lead to various impacts onto the acoustic modes:
  
\begin{itemize}
 \item In a first set of DNS, the instability behaves in a similar way
than observed in purely radiative simulations: there is a linear growth and a
nonlinear saturation of the acoustic oscillations. The bulk of kinetic
energy is embedded in acoustic motions, while convection participates
up to $20\%$ to the energy budget. However, on the contrary to what was observed
in Paper II, the nonlinear saturation does not involve any resonance between
modes. The frequential analysis emphasises numerous harmonics of the fundamental
acoustic mode that are probably indicative of a mono-mode saturation. In this
case, the convection does not affect much the oscillations.

 \item In a second category of DNS, the convective motions have a stronger
influence onto the acoustic oscillations. There is no clear nonlinear
saturation of the $\kappa$-instability while the amplitudes of 
acoustic modes remain very weak. Moreover, the kinetic energy is almost entirely
due to the downward-directed convective eddies. The vanishing of harmonics in
the frequential analysis is a further evidence of the weakness of the mode
amplitudes. In this case, convective plumes are quenching the oscillations and
this physical phenomenon looks like to what is expected to occur in the
coldest Cepheids close to the red edge of the instability strip. In these
stars, the large surface convective zone is indeed suspected to stabilise the
global radial oscillations. 

\end{itemize}

The various behaviours obtained in these local simulations of the
$\kappa$-mechanism with convection have been further studied in the last part
of this paper, where we have given some hints to explain the physical
conditions that lead to the mode quenching. Both the influence of the
amount of heat carried by convection and the density contrast across
the layer have been investigated. We have first shown that even in the
inefficient regime where convection only carries $10-15\%$
of the total heat flux, the convective plumes may cancel the radial
oscillations,
indicating that the strength of convection is not the only parameter that explains the
mode quenching. The density contrast seems on the contrary to play an important
role on the dynamics of acoustic modes. In fact, weaker stratifications (that
is, higher values of the density scale $H_{\rho}$) correspond to bigger
vortices. It means that the scale of the more powerful structure in the flow is
larger when the stratification is weaker. In the less stratified simulations,
where the acoustic modes appear to be strongly quenched by convection, the size
of convective plumes is indeed larger and that leads to an important horizontal
filling-factor. In our DNS, the amplitude of radial pulsations (and their
nonlinear saturation) may thus be governed by the screening effect of large
convective vortices.

The 2-D DNS of the $\kappa$-mechanism with convection presented in this
paper are a first step
in our study of the convection-pulsation coupling occurring in the
coldest Cepheid stars with an outer convective zone.
They are based on a simplified physical model that
keeps the main physics responsible for the excitation of acoustic modes by the
$\kappa$-mechanism. The opacity bump is for instance modelled by a
hollow in the 
radiative conductivity profile whom the shape is adjusted to get
unstable acoustic modes superimposed with convection. One of the main prospects of this pioneering
work is thus to improve this model toward more realistic stellar
conditions (e.g. inputs from tabulated opacities or coding of a free-surface 
boundary condition for the
velocity). Such developments
are however challenging as they will require huge numerical improvements in the
Pencil Code.


Another interesting prospect is the comparison between our 2-D DNS and
the prescriptions of different time-dependent convection (TDC) models
\citep[e.g.][]{St,K,YKB98}. In fact, even if our simplified model of the
$\kappa$-mechanism is far from the realistic stellar parameters, it could be
interesting to compare in our simulations the temporal evolution of the 
convective and kinetic fluxes modulated by acoustic oscillations
with their TDC counterparts. We will check the
validity of such approaches and give some constraints on the multiple
dimensionless $\alpha$ coefficients involved in these 1-D models.

\begin{acknowledgements}
This work was granted access to the HPC resources of CALMIP under the
allocation 2010-P1021. It is also a pleasure to thank J\'er\^ome Ballot for his
fruitful help on hydrostatic stellar models.
\end{acknowledgements}


\end{document}